\documentclass[aps,prd,10pt,twocolumn,nofootinbib,superscriptaddress]{revtex4-2}
\usepackage{epsfig}
\usepackage{multirow}
\usepackage{bm}
\usepackage{latexsym}
\usepackage{natbib}
\usepackage{url}
\usepackage{balance}
\usepackage[T1]{fontenc}
\usepackage{lmodern}

\usepackage{color}
\usepackage{amsfonts,amssymb,amsmath}
\usepackage{graphicx,epsfig}
\usepackage{psfrag}
\usepackage{float}
\usepackage{subfigure}
\usepackage{subcaption}
\usepackage[citecolor = ProcessBlue]{hyperref}
\hypersetup{colorlinks=true}
\usepackage{mathtools}
\usepackage{enumitem}
\usepackage{float}
\usepackage{mathrsfs}
\usepackage[dvipsnames]{xcolor}
\usepackage[utf8]{inputenc}
\usepackage{comment}
\allowdisplaybreaks
\definecolor{darkblue}{rgb}{0.0, 0.0, 0.62}
\definecolor{deepmagenta}{rgb}{0.7, 0.01, 0.7}
\definecolor{darkred}{rgb}{0.55, 0.0, 0.0}
\begin{document}

\title{Connecting quasi-normal modes with causality in Lovelock theories of gravity}
\author{Avijit Chowdhury}
\email{avijit.chowdhury@rnd.iitg.ac.in}
\affiliation{Indian Institute of Technology Guwahati, Assam-781039, India}
\author{Akash K Mishra}
\email{akash.mishra@icts.res.in}
\affiliation{International Centre for Theoretical Sciences, Tata Institute of Fundamental Research, Bangalore 560089, India}%
\affiliation{Theory Division, Saha Institute of Nuclear Physics,
1/AF Bidhan Nagar, Kolkata 700064, India}%
\author{Sumanta Chakraborty}
\email{tpsc@iacs.res.in}
\affiliation{School of Physical Sciences, Indian Association for the Cultivation of Science, Kolkata-700032, India}%
%%%%%%%%%%%%%%%%%%%%%%%%%%%%%%%%%%%%%%%%%%%%%%%%%%%%%%%%%%%%%%%%%%%%%%%%%%%%%%%%%%%
%%%%%%%%%%%%%%%%%%%%%%%%%%%%%%%%%%%%%%%%%%%%%%%%%%%%%%%%%%%%%%%%%%%%%%%%%%%%%%%%%%%
%%%%%%%%%%%%%%%%%%%%%%%%%%%%%%%%%%%%%%%%%%%%%%%%%%%%%%%%%%%%%%%%%%%%%%%%%%%%%%%%%%%
\begin{abstract}
The eikonal correspondence between the quasi-normal modes (QNMs) of asymptotically flat static spherically symmetric black holes and the properties of unstable null circular geodesics is studied in the case of higher dimensional Lovelock black holes (BHs). It is known that such correspondence does not generically hold for gravitational QNMs associated with BHs in Lovelock theories. In the present work, we revisit this correspondence and establish the relationship between the eikonal QNMs and the causal properties of the gravitational field equations in Lovelock theories of gravity.
\end{abstract}
%%%%%%%%%%%%%%%%%%%%%%%%%%%%%%%%%%%%%%%%%%%%%%%%%%%%%%%%%%%%%%%%%%%%%%%%%%%%%%%%%%%
%%%%%%%%%%%%%%%%%%%%%%%%%%%%%%%%%%%%%%%%%%%%%%%%%%%%%%%%%%%%%%%%%%%%%%%%%%%%%%%%%%%
%%%%%%%%%%%%%%%%%%%%%%%%%%%%%%%%%%%%%%%%%%%%%%%%%%%%%%%%%%%%%%%%%%%%%%%%%%%%%%%%%%%
\balance
\maketitle
\sloppy
%%%%%%%%%%%%%%%%%%%%%%%%%%%%%%%%%%%%%%%%%%%%%%%%%%%%%%%%%%%%%%%%%%%%%%%%%%%%%%%%%%%
%%%%%%%%%%%%%%%%%%%%%%%%%%%%%%%%%%%%%%%%%%%%%%%%%%%%%%%%%%%%%%%%%%%%%%%%%%%%%%%%%%%
%%%%%%%%%%%%%%%%%%%%%%%%%%%%%%%%%%%%%%%%%%%%%%%%%%%%%%%%%%%%%%%%%%%%%%%%%%%%%%%%%%%
\section{Introduction} \label{sec:1}

The study of black hole (BH) quasi-normal modes (QNMs) has been a topic of active research since the pioneering work of Vishveshwara~\cite{vishu1970PRD, vishu1970Nat, Leaver:1985ax, Berti:2018vdi, Berti:2005eb, Dias:2015wqa, London:2014cma}, and has gained significant interest from the direct detection of these modes in recent gravitational wave observations involving binary BH coalescences~\cite{LIGOScientific:2019fpa, Carullo:2019flw, Ghosh:2021mrv, Carullo:2021dui, Chakraborty:2017qve, Mishra:2023kng}. The QNM frequencies characterise the ringdown phase of the gravitational wave signal and are, in general, complex, encoding the characteristic parameters of the remnant BH formed due to the compact binary BH merger. Hence, `black hole spectroscopy' or the study of QNMs in the ringdown phase has emerged as a powerful tool in gravitational wave astronomy to study the nature of compact objects and to test deviations from general relativity~\cite{LIGOScientific:2016aoc, destounis2023arxiv, Mishra:2021waw}. The QNM frequencies are, in general, complex, owing to the fact that BHs describe a dissipative system, and the sign of the imaginary part of the QNM frequencies determines the stability of the BHs. Though the determination of the QNMs mostly requires numerical approaches, it is possible to obtain analytical expressions for the QNMs in the high-frequency (eikonal) limit. In this geometric optics regime, QNMs for BHs in general relativity are often interpreted as wave packets trapped at the unstable null circular geodesics, corresponding to photon sphere~\cite{cardoso2009PRD, konoplya2011RMP, berti2009CQG, kokkotas1999LRR} of the stationary spherically symmetric asymptotically flat BHs and slowly leaking out. Specifically, the real part of the QNM frequency is given by the angular velocity of the unstable null geodesics, whereas the imaginary part is determined by the principal Lyapunov exponent characterising the instability timescale of the orbit~\cite{ferrari1984PRD, mashhoon1985PRD, berti2009CQG},
%%%%%%%%%%%%%%%%%%%%%%%%%%%%%%%%%%%%%%%%%%%%%%%%%%%%%%%%%%%%%%
\begin{equation}\label{eq:eikonalWKB}
\omega_{n}=\Omega_{\rm ph} \ell - i \left(n+\frac{1}{2}\right) \mid\lambda_{\rm ph}\mid~,
\end{equation}
%%%%%%%%%%%%%%%%%%%%%%%%%%%%%%%%%%%%%%%%%%%%%%%%%%%%%%%%%%%%%%
where $n$ is the overtone number and one demands $\ell\gg n$. The above is an inherent property of BHs in general relativity and establishes a relation between the eikonal QNMs and the photon sphere~\cite{Jusufi:2020dhz, Yang:2021zqy, chen2023PLB,  Pedrotti:2024znu}. Whether such an eikonal correspondence is expected between the QNM frequencies and the unstable null geodesics in modified theories of gravity is a question worth addressing.

Among a large class of modified theories of gravity, the ones following Lovelock's theorem~\cite{lovelock1971JMP} give rise to second-order gravitational field equations and are free from Ostrogradsky instability. The theorem uniquely fixes Einstein's gravity in four spacetime dimensions. However, in $d>4$, the uniqueness of Einstein's gravity is lost, and there can be additional curvature-dependent terms that still allow for the equation of motion with, at most, second-order time derivatives. These are called Lovelock theories of gravity (see \cite{paddy2013PR} for an excellent review) and correspond to the immediate generalization of general relativity, involving a polynomial in powers of Riemann curvature. Given the special status of the Lovelock class of gravity theories, initial value formulation and causality for these theories has been of significant interest~\cite{choquet1988JMP, reall2014CQG, brustein2018PRD}.

In general, the causal structure of a gravity theory can be attributed to the properties of the characteristic hypersurfaces associated with the field equations~\cite{courant-hilbert}. For Einstein's theory, the characteristic hypersurfaces are necessarily null with respect to the background spacetime geometry. However, in Lovelock theories, the location of the characteristic hypersurfaces is determined by both the metric and the Riemann tensor, and hence, these hypersurfaces are in general non-null with respect to the background metric (referred to as the physical metric), suggesting sub(super)-luminal propagation of gravitational degrees of freedom. 

Interestingly, for BHs in Lovelock theories, these characteristic hypersurfaces can be made null, but not with respect to the physical metric, rather with respect to certain specific effective metrics. These effective metrics need to be Lorentzian for the theory to be hyperbolic, which is true for large BHs (these are BHs having mass large compared to the coupling strength), whereas for small BHs (the mass of the BH is small or comparable to the coupling strength), these effective metrics may change signature outside the horizon, which makes the theory non-hyperbolic. Thus, in order to have an appropriate initial value formulation, one must limit the size of a BH in Lovelock theories of gravity (or, equivalently, the `size' of the Lovelock corrections). This leads to the consensus that Lovelock theories can only be considered as perturbative corrections to Einstein's theory in higher dimensions~\cite{papallo2015JHEP, brustein2018PRD}. On the other hand, these problems are avoided if one considers a single term in the Lovelock polynomial, which is referred to as a pure Lovelock theory of gravity. The pure Lovelock theories are characterized by their order $N$, implying that the Lagrangian for the theory consists of $N$-th power of the Riemann tensor. Though this theory cannot be considered as a correction to general relativity, its characteristic hypersurfaces are also not null in the physical metric describing BH spacetimes in pure Lovelock theories~\cite{dadhich2016PRD, Chakraborty:2015kva, Chakraborty:2016qbw, Malafarina:2020pvl, Gannouji:2019gnb}.   

The fact that different graviton polarizations propagate along the null direction of effective metrics, together with the observation that the eikonal gravitational QNMs in the Lovelock theories are unrelated to the properties of the unstable null circular geodesics of the physical metric~\cite{konoplya2017PLB,Moura:2021eln}, lead us to enquire whether the eikonal gravitational QNMs are instead related to the unstable null circular geodesics of the effective metric (the unstable null circular geodesics of the effective metric in the dynamical context has been discussed in \cite{mishra2019PRD}). Such a correspondence, though expected~\cite{glampedakis2019PRD}, has never been explored in detail to the best of our knowledge. This work is an attempt to fill this gap in the literature and provide a direct physical interpretation of the eikonal gravitational QNMs in Lovelock theories of gravity in terms of the geodesic properties of the effective metric and thus relate the eikonal QNMs with the causality and hyperbolicity of these theories. In this paper, we concentrate on (i) quadratic Einstein-Gauss-Bonnet theory in $d>4$ dimension as a working example of higher dimensional Lovelock theory and (ii) Pure Lovelock gravity of the $N$-th order.

The paper is organised as follows: in section~\ref{sec:2A}, we discuss static spherically symmetric as well as asymptotically flat BHs in general Lovelock theories of gravity (see section~\ref{sec:2Ai}) and pure Lovelock gravity (see section~\ref{sec:2Aii}). Subsequently, we review the effective metrics for different propagating modes and their connection with the physical metric in section~\ref{sec:2B}, with the case of Einstein-Gauss-Bonnet gravity discussed in section~\ref{sec:2Bi} and the pure Lovelock gravity in section~\ref{sec:2Bii}. In section~\ref{sec:3}, we review the eikonal QNM frequencies associated with gravitational perturbations of the Einstein-Gauss-Bonnet and pure Lovelock BHs using WKB approximation and establish their correspondence with the Lyapunov exponent and the angular velocity of the null geodesics of the effective metric in section ~\ref{sec:2B}. To further validate this analogy, section \ref{sec:4} compares the QNM frequencies evaluated numerically with the ones obtained by eikonal approximation. Finally, we conclude in section~\ref{sec:5} with a brief discussion of our results and their significance. Some additional computations have been presented in Appendix \ref{AppA}. 

%%%%%%%%%%%%%%%%%%%%%%%%%%%%%%%%%%%%%%%%%%%%%%%%%%%%%%%%%%%%%%%%%%%%%%%%%%%%%%%%%%%
%%%%%%%%%%%%%%%%%%%%%%%%%%%%%%%%%%%%%%%%%%%%%%%%%%%%%%%%%%%%%%%%%%%%%%%%%%%%%%%%%%%
%%%%%%%%%%%%%%%%%%%%%%%%%%%%%%%%%%%%%%%%%%%%%%%%%%%%%%%%%%%%%%%%%%%%%%%%%%%%%%%%%%%
%%%%%%%%%%%%% sec:2 %%%%%%%%%
%%%%%%%%%%%%%%%%%%%%%%%%%%%%%%%%%%%%%%%%%%%%%%%%%%%%%%%%%%%%%%%%%%%%%%%%%%%%%%%%%%%
%%%%%%%%%%%%%%%%%%%%%%%%%%%%%%%%%%%%%%%%%%%%%%%%%%%%%%%%%%%%%%%%%%%%%%%%%%%%%%%%%%%
%%%%%%%%%%%%%%%%%%%%%%%%%%%%%%%%%%%%%%%%%%%%%%%%%%%%%%%%%%%%%%%%%%%%%%%%%%%%%%%%%%%
\section{Causality in Lovelock theories: Effective metric}\label{sec:2}

Lovelock theories of gravity correspond to a unique class of higher curvature gravity theories in higher spacetime dimensions having second order dynamical equations. In four dimensions, general relativity is the unique Lovelock theory, while in higher dimensions, there can be other higher curvature contributions. For example, in spacetime dimensions $d\geq 5$, the Lovelock theory, besides the Einstein term also involves the Gauss-Bonnet contribution $\mathcal{G}=R^{2}-4R_{ab}R^{ab}+R_{abcd}R^{abcd}$. Note that, only the above specific combination of quadratic curvature term yields second order field equations for the metric. In a spacetime with dimensions $d$, the Lovelock theory is a finite series in the powers of the curvature till a term $(\textrm{Riemann})^{N_{\rm max}}$, where $2N_{\rm max}+1\leq d$. Besides the general Lovelock polynomial, one can also work with a single higher curvature term in the Lovelock series, known as pure Lovelock theories. For example, the Gauss-Bonnet term alone depicts a pure Lovelock theory of gravity. Both of these theories have interesting features, e.g., they have second order field equations and hence free from ghost-like instabilities \cite{Woodard:2015zca}, they satisfy the laws of black hole mechanics, as well as thermodynamics \cite{Chakraborty:2015wma, Chakraborty:2014rga, Chakraborty:2014joa, Chakraborty:2018dvi}, and yield a well-posed initial value problem \cite{papallo2015JHEP} under specific conditions. However, the Lovelock theories of gravity have one peculiar feature, unlike general relativity, where both the independent gravitational degrees of freedom propagate at the speed of light, in Lovelock theories of gravity the propagation speeds of the independent gravitational degrees of freedom are different. We elaborate on this point, which is the central theme of this work, related to causality in Lovelock theories of gravity, below. 

%%%%%%%%%%%%%%%%%%%%%%%%%%%%%%%%%%%%%%%%%%%%%%%%%%%%%%%%%%%%%%%%%%%%
%%%%%%%%%%%%%%%%%%%%%%%%%%%%%%%%%%%%%%%%%%%%%%%%%%%%%%%%%%%%%%%%%%%%
%%%%%%%%%%%%%%%%%%%%%%%%%%%%%%%%%%%%%%%%%%%%%%%%%%%%%%%%%%%%%%%%%%%%
\subsection{Effective metric for gravitational perturbation in Lovelock gravity}\label{sec:2A}

In this section, we introduce the idea of an effective metric for gravitational degrees of freedom associated with generic Lovelock theories of gravity in higher spacetime dimensions. We start by spelling out the Lagrangian density for a general Lovelock theory in $d$ spacetime dimensions, which read~\cite{lovelock1971JMP, reall2014CQG}, 
%%%%%%%%%%%%%%%%%%%%%%%%%%%%%%%%%%%%%%%%%%%%%%%%%%%%%%%%%%%%%%
\begin{align}\label{eq:lovelock-L}
\mathcal{A}&=\frac{1}{16\pi}\int d^{d}x\,\mathscr{L}\,,
\nonumber
\\
\mathscr{L}&=\sum_{p=1}^{N_{\rm max}} 2^{-p}\alpha_{p} \delta^{c_1\ldots c_{2p}}_{d_1\ldots d_{2p}} R_{c_1 c_2}^{d_1 d_2} \cdots R_{c_{2p-1} c_{2p}}^{d_{2p-1} d_{2p}}~,
\end{align}%notation using Reall Tanahashi
%%%%%%%%%%%%%%%%%%%%%%%%%%%%%%%%%%%%%%%%%%%%%%%%%%%%%%%%%%%%%%
where, $N_{\rm max}\leq \{(d-1)/2\}$. The coupling constants $\alpha_{p}$ are usually dimensionful and different for different powers of the curvature, with $\alpha_{1}=1$, where the Newton's gravitational constant has been set to unity, and hence effectively all of the other coupling constants are scaled by the Newton's constant. The tensors $R_{ab}^{cd}$ are the background Riemann tensors, and $\delta^{ab\cdots}_{cd\cdots}$ is the determinant tensor, which is completely anti-symmetric in both the upper and lower indices and has the same symmetry properties as the Riemann tensor in each set of four indices. Our interest lies in the static and spherically symmetric black hole solutions, which are described by the following metric ansatz, 
%%%%%%%%%%%%%%%%%%%%%%%%%%%%%%%%%%%%%%%%%%%%%%%%%%%%%%%%%%%%%%
\begin{equation}\label{eq:phy-metric}
ds^2=-f(r) dt^2+f(r)^{-1} dr^2 +r^2 d\Omega^2_{d-2}~,
\end{equation}
%%%%%%%%%%%%%%%%%%%%%%%%%%%%%%%%%%%%%%%%%%%%%%%%%%%%%%%%%%%%%%
where $d\Omega^2_{d-2}$ is the metric on the $(d-2)$ dimensional unit sphere. It is instructive to define the function, $\psi(r)\equiv r^{-2}\{1-f(r)\}$, which in Lovelock theories of gravity satisfies the following algebraic relation,
%%%%%%%%%%%%%%%%%%%%%%%%%%%%%%%%%%%%%%%%%%%%%%%%%%%%%%%%%%%%%%
\begin{equation}\label{eq:W}
W[\psi(r)]\equiv \psi+\sum_{p=2}^{N_{\rm max}}\Big[\alpha_{p}\prod_{k=1}^{2p-2}(d-2-k)\psi^{p}\Big]=\frac{\mu}{r^{d-1}}~,
\end{equation}
%%%%%%%%%%%%%%%%%%%%%%%%%%%%%%%%%%%%%%%%%%%%%%%%%%%%%%%%%%%%%%
where $\mu$ is related to the ADM mass $M$ of the static and spherically symmetric black hole spacetime as, 
%%%%%%%%%%%%%%%%%%%%%%%%%%%%%%%%%%%%%%%%%%%%%%%%%%%%%%%%%%%%%%
\begin{equation}
\mu=\frac{4\Gamma\left(\frac{d-1}{2}\right)M}{\pi^{\frac{d-3}{2}}}~;
\quad
M\equiv S_{d-2}\int dr\,r^{d-2}\left(-T^{t}_{t}\right)~,
\end{equation}
%%%%%%%%%%%%%%%%%%%%%%%%%%%%%%%%%%%%%%%%%%%%%%%%%%%%%%%%%%%%%%
where, $S_{d-2}$ depicts the surface area of the $(d-2)$ sphere. Note that for $d=4$, it follows that $N_{\rm max}=1$, and hence we obtain $\mu=2M$, consistent with the Schwarzschild solution in general relativity. 

Note that both massless scalars as well as photons travel with the speed of light and execute circular motion at a radius $r_{\rm ph}$, satisfying $2f(r_{\rm ph})=r_{\rm ph}f'(r_{\rm ph})$. As a consequence, the real and imaginary parts of eikonal QNMs associated with massless scalars and photons are given by the angular velocity of photons and Lyapunov exponent associated with the photon circular orbit, located at $r_{\rm ph}$, respectively. However, for gravitational perturbation, the situation is different, as the gravitational degrees of freedom travel with different velocities and hence do not seem to have any direct correspondence with the physics at radius $r_{\rm ph}$ \cite{choquet1988JMP}. 

In order to find out any connection of the eikonal QNMs associated with gravitational perturbation with the effective metric experienced by the gravitational perturbations, we consider the decomposition of the gravitational perturbations around the spherically symmetric background spacetime into scalar ($S$), vector ($V$) and tensor ($T$) types. Note that the tensor type perturbation is absent in general relativity, while the scalar type perturbation connects to Zerilli and the vector type perturbation corresponds to Regge-Wheeler. While in Einstein gravity, the Zerilli and the Regge-Wheller perturbations are related to each other and propagate with the same speed, in higher dimensional Lovelock gravity, each type of perturbation propagates with different speeds and `sees' a different effective metric, which has the following general form,
%%%%%%%%%%%%%%%%%%%%%%%%%%%%%%%%%%%%%%%%%%%%%%%%%%%%%%%%%%%%%%
\begin{equation}\label{eq:eff-metric-LL}
G_{ab}^{A} dx^{a} dx^{b}= -f(r) dt^2 + f(r)^{-1} dr^2+\frac{r^2}{c_A (r)} d\Omega^2_{d-2}~,
\end{equation}
%%%%%%%%%%%%%%%%%%%%%%%%%%%%%%%%%%%%%%%%%%%%%%%%%%%%%%%%%%%%%%
where the superscript $A \in \{0,S,V,T\}$, with the choice $A=0$ signifying the physical metric as in Eq. \eqref{eq:phy-metric} with $c_0(r)=1$. For other choices of $A$, the signature of $c_{A}(r)$ determines the Lorentzian character of the metric. For example, the inverse of the effective metric\footnote{Note that the metric $G_{A}^{ab}$ is defined as the inverse of $G^{A}_{ab}$ and \emph{not} by raising the indices on $G^{A}_{ab}$ using the physical metric.} $G_{A}^{ab}$ is degenerate if $c_{A}(r)=0$, Lorentzian if $c_{A}(r)>0$, while $c_{A}(r)<0$ suggests a breakdown of the hyperbolicity, whereas $c_{A}(r)>1$ implies superluminal propagation. Following~\cite{reall2014CQG}, we present below the expressions for $c_{A}(r)$ for scalar, vector, and tensor perturbations of black holes in the Lovelock theories of gravity, which become
%%%%%%%%%%%%%%%%%%%%%%%%%%%%%%%%%%%%%%%%%%%%%%%%%%%%%%%%%%%%%%
\begin{align}\label{eq:cA}
c_S(r)=&3\Bigl(1-\frac{1}{d-2}\Bigr)\mathscr{A}(r)+\Bigl(1-\frac{3}{d-2}\Bigr)\frac{1}{\mathscr{A}(r)}
\nonumber
\\
&-\Bigl(1-\frac{2}{d-2}\Bigr)(\mathscr{B}(r)+3)~,
\\
c_V(r)=&\mathscr{A}(r)~,
\\
c_T(r)=&-\Bigl(1+\frac{1}{d-4}\Bigr)\mathscr{A}(r)-\Bigl(1-\frac{1}{d-4}\Bigr)\frac{1}{\mathscr{A}(r)}
\nonumber
\\
&+\mathscr{B}(r)+3~.
\end{align}
%%%%%%%%%%%%%%%%%%%%%%%%%%%%%%%%%%%%%%%%%%%%%%%%%%%%%%%%%%%%%%
In the above, we have introduced two radial functions $\mathscr{A}(r)$ and $\mathscr{B}(r)$, having the following form,
%%%%%%%%%%%%%%%%%%%%%%%%%%%%%%%%%%%%%%%%%%%%%%%%%%%%%%%%%%%%%%
\begin{align}
\mathscr{A}(r)&=1-\frac{d-1}{d-3}\frac{W \partial^2_{\psi}W}{(\partial_\psi W)^2}~,
\\
\mathscr{B}(r)&=\frac{(d-1)^2}{(d-3)(d-4)}\frac{W^2 \partial^3_{\psi}W}{\mathscr{A}(r)(\partial_\psi W)^3}~,
\label{eq:AB}
\end{align}
%%%%%%%%%%%%%%%%%%%%%%%%%%%%%%%%%%%%%%%%%%%%%%%%%%%%%%%%%%%%%%
where, $W[\psi]$ has been introduced in Eq.\eqref{eq:W}. It is worth emphasizing that in~\cite{brustein2018PRD}, the authors derived the effective metric using a slightly different scheme, which differs from Eq.~\eqref{eq:eff-metric-LL} only by an overall conformal factor and hence does not affect the properties of the null geodesics, which we are interested in. 
%%%%%%%%%%%%%%%%%%%%%%%%%%%%%%%%%%%%%%%%%%%%%%%%%%%%%%%%%%%%%%%%%%%%
%%%%%%%%%%%%%%%%%%%%%%%%%%%%%%%%%%%%%%%%%%%%%%%%%%%%%%%%%%%%%%%%%%%%
%%%%%%%%%%%%%%%%%%%%%%%%%%%%%%%%%%%%%%%%%%%%%%%%%%%%%%%%%%%%%%%%%%%%
\subsubsection{Einstein-Gauss-Bonnet gravity}\label{sec:2Ai}

As we have described above, the effective metric experienced by various gravitational degrees of freedom in Lovelock theories of gravity are given by Eq.\eqref{eq:eff-metric-LL}, with the metric functions $c_{A}(r)$, given by Eqs.\eqref{eq:cA}---\eqref{eq:AB}. Determining the function $f(r)$, on the other hand, appearing in the effective metric for Lovelock gravity, is a difficult task. Thus, we concentrate on the first non-trivial correction to the Einstein-Hilbert action, namely the Gauss-Bonnet term. The function $f(r)$ in the context of $d$ dimensional vacuum spacetime in Einstein-Gauss-Bonnet gravity, takes the following form,
%%%%%%%%%%%%%%%%%%%%%%%%%%%%%%%%%%%%%%%%%%%%%%%%%%%%%%%%%%%%%%
\begin{equation}\label{eq:5d-GB-gtt}
f(r)=1+\frac{r^2}{2\widetilde{\alpha}_{2}}\Big[1-q(r)\Big]\,, 
\quad 
q(r)\equiv\sqrt{1+\frac{4\widetilde{\alpha}_{2}\,\mu}{r^{d-1}}}\,,
\end{equation}
%%%%%%%%%%%%%%%%%%%%%%%%%%%%%%%%%%%%%%%%%%%%%%%%%%%%%%%%%%%%%%
where $\widetilde{\alpha}_{2}\equiv (d-3)(d-4)\alpha_{2}$, with $\alpha_{2}$ representing the Gauss-Bonnet coupling constant which measures the deviation from general relativity ($\alpha_{2}=0$ corresponds to general relativity). 
For all subsequent numerical computations, we set the mass parameter $\mu = 1$.
In what follows, we will use the Einstein-Gauss-Bonnet gravity as a proxy for general Lovelock theories, as it captures all the basic ingredients associated with Lovelock theories of gravity.  

%%%%%%%%%%%%%%%%%%%%%%%%%%%%%%%%%%%%%%%%%%%%%%%%%%%%%%%%%%%%%%%%%%%%
%%%%%%%%%%%%%%%%%%%%%%%%%%%%%%%%%%%%%%%%%%%%%%%%%%%%%%%%%%%%%%%%%%%%
%%%%%%%%%%%%%%%%%%%%%%%%%%%%%%%%%%%%%%%%%%%%%%%%%%%%%%%%%%%%%%%%%%%%
\subsubsection{Pure Lovelock gravity}\label{sec:2Aii}

Rather than considering the full Lovelock polynomial, one can consider a single term in the Lovelock series in a given spacetime dimension, which is referred to as the pure Lovelock theory of gravity. The pure Lovelock theories are characterized by two quantities: the spacetime dimension $d$ and the order $N$ of the Riemann tensor present in a specific Lovelock term. From our previous discussion and also \cite{dadhich2013PRD}, it follows that such a Lovelock term will admit nontrivial vacuum solutions only if $d\geq 2N+2$, as in $d=2N+1$, a pure Lovelock theory of order $N$ is purely topological~\cite{dadhich2012PLB}. Moreover, the stability of black holes in pure Lovelock gravity holes has also been studied in detail in \cite{gannouji2014CQG, gannouji2019PRD}, and the black holes are found to be stable in $d\geq3N+1$ dimensions.

A pure Lovelock theory of order $N$ admits static spherically symmetric black hole solutions of the form given in Eq.\eqref{eq:phy-metric}, with the radial function $f(r)$ taking the following form,
%%%%%%%%%%%%%%%%%%%%%%%%%%%%%%%%%%%%%%%%%%%%%%%%%%%%%%%%%%%%%%
\begin{equation}\label{eq:pLL-phys-metric}
f(r)=1-\left(\frac{r_{+}}{r}\right)^{\frac{d-2N-1}{N}}~,
\end{equation}
%%%%%%%%%%%%%%%%%%%%%%%%%%%%%%%%%%%%%%%%%%%%%%%%%%%%%%%%%%%%%%
where $r_{+}$ is the location of the event horizon~\cite{dadhich2011MT, cai2006PRD, cai2008PRD}. Given the above expression for $f(r)$, it follows that $\psi(r)=r_{+}^{-2}(r_{+}/r)^{(d-1)/N}$, and hence the quantity $W[\psi]$, defined in Eq.~\eqref{eq:W}, for $N$ order pure Lovelock gravity simplifies to,
%%%%%%%%%%%%%%%%%%%%%%%%%%%%%%%%%%%%%%%%%%%%%%%%%%%%%%%%%%%%%%
\begin{align}\label{eq:WPLL}
W[\psi(r)]&=\frac{\alpha_{N}(d-3)!}{(d-2N-1)!}\psi^{N}=\frac{\mu}{r^{d-1}}~,
\\
\mu&=\left(\frac{\alpha_{N}(d-3)!}{(d-2N-1)!}\right)r_{+}^{d-2N-1}~,
\end{align}
%%%%%%%%%%%%%%%%%%%%%%%%%%%%%%%%%%%%%%%%%%%%%%%%%%%%%%%%%%%%%%
which is obtained by replacing the summation on the left hand side of Eq.\eqref{eq:W} by a single term. Moreover, the above exercise directly provides the correspondence between the parameter $\mu$, the horizon radius $r_{+}$ and the coupling $\alpha_{N}$. This results in relatively simple expressions for the functions $c_A(r)$, in the context of pure Lovelock theories of gravity:
%%%%%%%%%%%%%%%%%%%%%%%%%%%%%%%%%%%%%%%%%%%%%%%%%%%%%%%%%%%%%%
\begin{eqnarray}
c_S(r)=\frac{d-1-N}{(d-2)N}~,
\label{eq:cA-PLL-1}
\\
c_V(r)=\frac{d-1-2N}{(d-3)N}~,
\label{eq:cA-PLL-2}
\\
c_T(r)=\frac{d-1-3N}{(d-4)N}~.
\label{eq:cA-PLL-3}
\end{eqnarray}
%%%%%%%%%%%%%%%%%%%%%%%%%%%%%%%%%%%%%%%%%%%%%%%%%%%%%%%%%%%%%%
Interestingly, in the critical dimension, $d=3N+1$, the spacetime metric of the pure Lovelock theories of gravity, reduces to the Schwarzschild solution and $c_T$ vanishes identically. This implies that in the above critical dimension, the tensor perturbations do not propagate. However, unlike general relativity, in this critical dimension, the effective metric for the scalar and the vector degrees of freedom are different. Thus, the effective metric associated with scalar and vector perturbations differs between general relativity and Lovelock gravity in $d=3N+1$ dimensions. As we will depict next, this will have implications for gravitational QNMs of pure Lovelock black holes. 

%%%%%%%%%%%%%%%%%%%%%%%%%%%%%%%%%%%%%%%%%%%%%%%%%%%%%%%%%%%%%%%%%%%%
%%%%%%%%%%%%%%%%%%%%%%%%%%%%%%%%%%%%%%%%%%%%%%%%%%%%%%%%%%%%%%%%%%%%
%%%%%%%%%%%%%%%%%%%%%%%%%%%%%%%%%%%%%%%%%%%%%%%%%%%%%%%%%%%%%%%%%%%%
\subsection{Gravitonsphere and its properties}\label{sec:2B}

Just as the circular geodesic of a photon, in a static and spherically symmetric background, is referred to as a photon sphere. In a similar spirit, the circular geodesic of the graviton (essentially gravitational wave in the high frequency/eikonal limit) in a static and spherically symmetric spacetime is referred to as the graviton sphere \cite{papallo2015JHEP}. These also refer to the unstable null circular geodesics of the spacetime described by the effective metric, which are distinct for scalar, vector and tensor perturbations, referred to as the scalar-sphere, vector-sphere and tensor-sphere, respectively. The location of the graviton sphere will be different for scalar, vector and tensor type perturbations and will be determined by the effective metric $G^{A}_{ab}$. Following the corresponding situation for the physical metric, null geodesics of the effective metric $G_{Aab}$ satisfy the following equation,
%%%%%%%%%%%%%%%%%%%%%%%%%%%%%%%%%%%%%%%%%%%%%%%%%%%%%%%%%%%%%%
\begin{equation}
\dot{r}^2+f(r) c_A(r)\frac{L^2}{r^{2}}=E^{2}~,
\end{equation}
%%%%%%%%%%%%%%%%%%%%%%%%%%%%%%%%%%%%%%%%%%%%%%%%%%%%%%%%%%%%%%
where $E$ and $L$ are the conserved energy and angular momentum of the graviton associated with the Killing fields $(\partial/\partial t)$ and $(\partial/\partial \phi)$, and are defined as
%%%%%%%%%%%%%%%%%%%%%%%%%%%%%%%%%%%%%%%%%%%%%%%%%%%%%%%%%%%%%%
\begin{equation}\label{eq:EL}
E=f(r)\dot{t}~, \qquad L=\frac{r^2}{c_A(r)}\dot{\phi}~.
\end{equation}
%%%%%%%%%%%%%%%%%%%%%%%%%%%%%%%%%%%%%%%%%%%%%%%%%%%%%%%%%%%%%%
The dot in Eq.~\eqref{eq:EL} denotes derivative with respect to the affine parameter and the effective potential $V_{\rm eff}^{A}$, that these null geodesics experience is given by, 
%%%%%%%%%%%%%%%%%%%%%%%%%%%%%%%%%%%%%%%%%%%%%%%%%%%%%%%%%%%%%%
\begin{equation}
V_{\rm eff}^{A}\equiv f(r)c_{A}(r)\left(\frac{L^{2}}{r^{2}}\right)~.
\end{equation}
%%%%%%%%%%%%%%%%%%%%%%%%%%%%%%%%%%%%%%%%%%%%%%%%%%%%%%%%%%%%%%
Alike the photon sphere, whether the effective potentials associated with scalar, vector and tensor perturbations have a single maxima is important for the subsequent analysis, since these correspond to the location of the unstable circular null geodesics associated with each graviton polarization. The location of these extrema, among the maxima, corresponds to the location of the graviton spheres $r^{A}_{\rm grav}$, can be determined by solving the following equation, 
%%%%%%%%%%%%%%%%%%%%%%%%%%%%%%%%%%%%%%%%%%%%%%%%%%%%%%%%%%%%%%
\begin{equation}
2f(r)c_A(r)=rf'(r)c_A(r)+rf(r)c_{A}'(r)~.
\end{equation}
%%%%%%%%%%%%%%%%%%%%%%%%%%%%%%%%%%%%%%%%%%%%%%%%%%%%%%%%%%%%%%
Note that for $A=0$, we have $c_{0}(r)=1$, and hence the above equation reduces to that of the photon sphere. In what follows, we will explicitly determine the location of the extrema and, hence, that of the gravitonspheres as well as the properties of the gravitonspheres in both Einsten-Gauss-Bonnet theory (as a proxy for general Lovelock models) and pure Lovelock theories of gravity.

%%%%%%%%%%%%%%%%%%%%%%%%%%%%%%%%%%%%%%%%%%%%%%%%%%%%%%%%%%%%%%%%%%%%
%%%%%%%%%%%%%%%%%%%%%%%%%%%%%%%%%%%%%%%%%%%%%%%%%%%%%%%%%%%%%%%%%%%%
%%%%%%%%%%%%%%%%%%%%%%%%%%%%%%%%%%%%%%%%%%%%%%%%%%%%%%%%%%%%%%%%%%%%
\subsubsection{Einstein-Gauss-Bonnet theory}\label{sec:2Bi}

%%%%%%%%%%%%%%%%%%%%
%%%%%%%%%%%%%%%%%%%%
%%%%%%%%%%%%%%%%%%%%
%%%%%%%%%%%%%%%%%%%%
\begin{figure*}[!tb]
\centering
\includegraphics[width=0.32\textwidth]{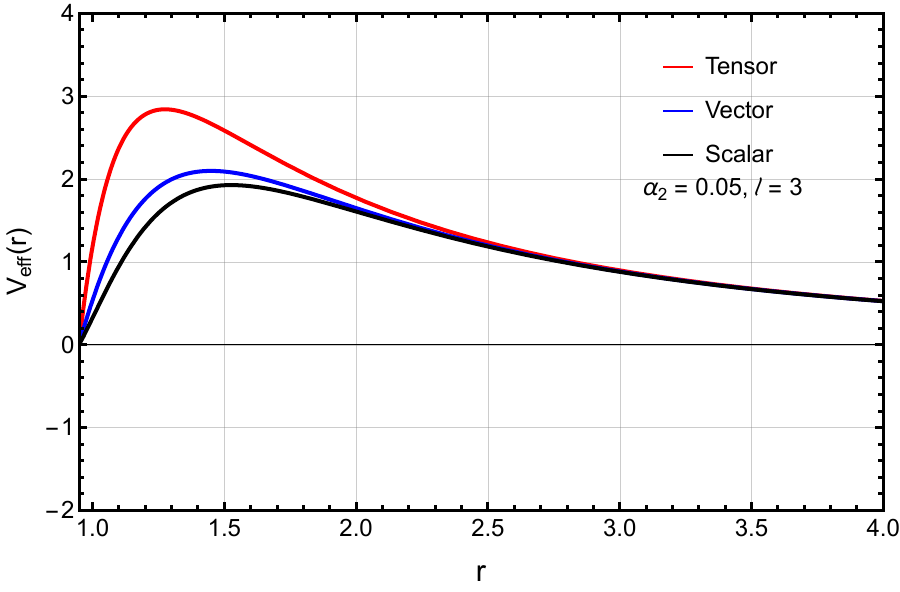}
\includegraphics[width=0.32\textwidth]{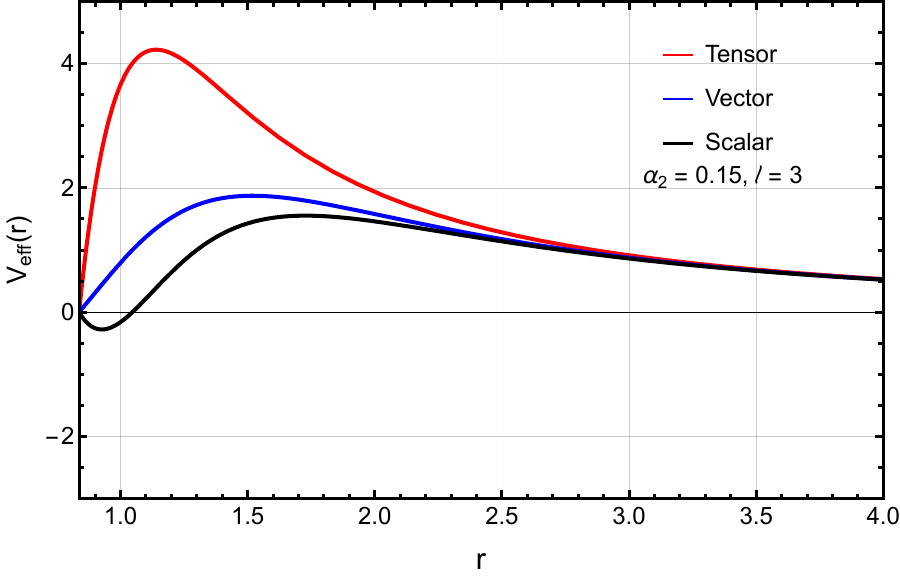}
\includegraphics[width=0.32\textwidth]{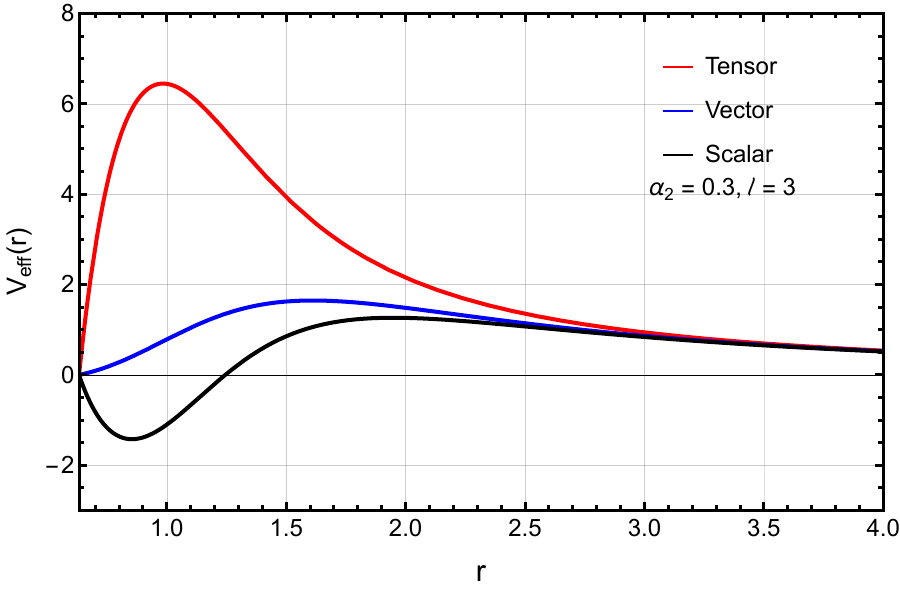}
\includegraphics[width=0.32\textwidth]{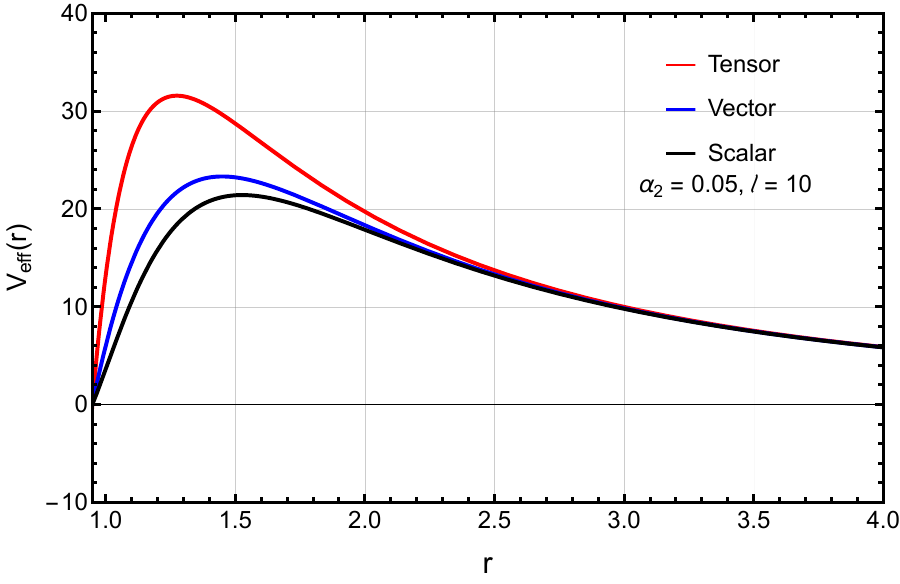}
\includegraphics[width=0.32\textwidth]{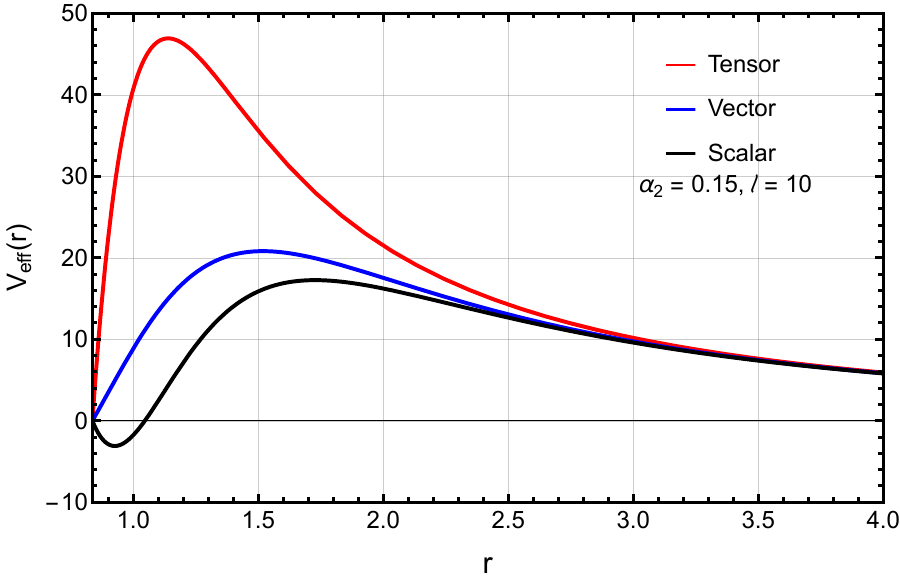}
\includegraphics[width=0.32\textwidth]{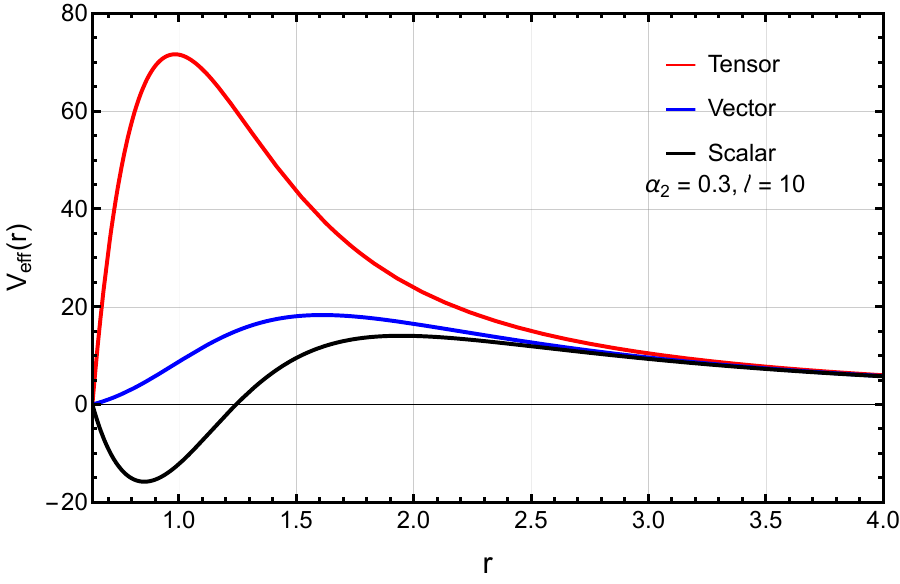}
\caption{
The effective potential for tensor, vector, and scalar modes of gravitational perturbations in five-dimensional Einstein-Gauss-Bonnet gravity for $\ell=3$ (top) and $\ell=10$ (bottom) for different values of the Gauss-Bonnet coupling parameter $\alpha_2$. The range of the x-axis begins at the horizon location determined by the corresponding value of $\alpha_2$.}
\label{fig:potential}
\end{figure*}
%%%%%%%%%%%%%%%%%%%%
%%%%%%%%%%%%%%%%%%%%
%%%%%%%%%%%%%%%%%%%%
%%%%%%%%%%%%%%%%%%%%

In this section, we will discuss the structure of the effective potential associated with Einstein-Gauss-Bonnet's gravity. As we will demonstrate, the structure of the effective potential will reveal that black holes in Lovelock theories of gravity are, in general, unstable for large values of the coupling parameters $\alpha_{m}$~\cite{gleiser2005PRD,dotti2005PRD, beroiz2007PRD, cuyubamba2016PRD, takahashi2010PTP1, konoplya2017JCAP}, associated with the existence of stable null geodesics, and hence minima of the effective potential. Following this, we have plotted the effective potential associated with all three types of gravitational perturbation in Fig. \ref{fig:potential} for two different choices of $\ell$, namely $\ell=3$ and $\ell=10$. As evident, the effective potentials associated with vector as well as tensor perturbation depict a single extremum, which is a maximum, irrespective of the value of the Gauss-Bonnet coupling parameter $\alpha_{2}$. While the effective potential for the scalar perturbation has a single maxima for smaller values of $\alpha$, but for larger values of the same it develops a minima as well, signalling instability. Thus, the effective potential approach, as considered here, connects naturally with previous literature and attributes the instability in the scalar sector of gravitational perturbation to the emergence of stable circular null geodesics in the graviton metric. In general, one can show that for small $\alpha_{2}$, i.e., if one ignores all terms of $\mathcal{O}(\alpha_{2}^{2})$, there will be one real root for the graviton metric and that will be a maximum. Whenever the Gauss-Bonnet parameter becomes large so that the above approximation breaks down, it starts to have more than one real root. This is precisely the phenomenon depicted in Fig. \ref{fig:potential}, as presented above. 

This allows us to restrict our attention to the small coupling parameter regime, and expand all the physical quantities in powers of the coupling parameter. The first is to compare the location of the photon and graviton spheres. It turns out that for $\alpha_{2}=0$, the photon and graviton sphere coincide, as expected, and hence, the radius of the gravitonspheres differ from that of the photonsphere (corresponding to $c_0=1$) at the leading order in the Gauss-Bonnet coupling constant, such that,
%%%%%%%%%%%%%%%%%%%%%%%%%%%%%%%%%%%%%%%%%%%%%%%%%%%%%%%%%%%%%%
\begin{align}
\Delta r^{T}_{\rm grav}&\equiv r^{T}_{\rm grav}-r_{\rm ph}
=-3\alpha_{2}\left(\frac{d-2}{d-4}\right)r_{\textrm{ph}(1)}~,
\label{eq:DrT}
\\
\Delta r^{V}_{\rm grav}&\equiv r^{V}_{\rm grav}-r_{\rm ph}
=\alpha_{2}\left(d-2\right)r_{\textrm{ph}(1)}~,
\label{eq:DrV}
\\
\Delta r^{S}_{\rm grav}&\equiv r^{S}_{\rm grav}-r_{\rm ph}
=(d-1)\alpha_{2}r_{\textrm{ph}(1)}~,
\label{eq:DrS}
\end{align}
%%%%%%%%%%%%%%%%%%%%%%%%%%%%%%%%%%%%%%%%%%%%%%%%%%%%%%%%%%%%%%
where, we have introduced $r_{\textrm{ph}(1)}$, the coefficient of $\alpha_{2}$ in the expansion of the photon sphere in Gauss-Bonnet coupling constant, which reads,
%%%%%%%%%%%%%%%%%%%%%%%%%%%%%%%%%%%%%%%%%%%%%%%%%%%%%%%%%%%%%%
\begin{equation}
r_{\textrm{ph}(1)}=-\frac{(d-4)\mu}{\bigl[\frac{(d-1)\mu}{2}\bigr]^{\frac{d-2}{d-3}}}~.
\label{eq:rc10}
\end{equation}
%%%%%%%%%%%%%%%%%%%%%%%%%%%%%%%%%%%%%%%%%%%%%%%%%%%%%%%%%%%%%%
Therefore, the difference between the location of the gravitonsphere from the photon sphere appears at linear order in the Gauss-Bonnet coupling parameter. The detailed expressions, involving both zeroth order and linear order terms in the Gauss-Bonnet coupling, for the graviton as well as the photon sphere are given in Appendix~\ref{AppA}. 

The next important property associated with the maxima of the effective potential is the angular velocity of a null geodesic at that maxima. These angular velocities at the gravitonspheres are given by,
%%%%%%%%%%%%%%%%%%%%%%%%%%%%%%%%%%%%%%%%%%%%%%%%%%%%%%%%%%%%%%
\begin{equation}\label{eq:omega-gen}
\Omega_{\rm grav}^{A} =\frac{\sqrt{c_{A}(r_{\rm grav}^{A})f(r_{\rm grav}^{A})}}{r_{\rm grav}^{A}}~,
\end{equation}
%%%%%%%%%%%%%%%%%%%%%%%%%%%%%%%%%%%%%%%%%%%%%%%%%%%%%%%%%%%%%%
where, $A$ denotes scalar, vector and tensor perturbations, collectively. Alike the location of the gravitonsphere, we can also expand the coordinate angular velocity in powers of the Gauss-Bonnet coupling $\alpha_{2}$. In this case as well the angular velocity of the graviton and photon spheres differ at the leading order in the Gauss-Bonnet Coupling constant, such that,
%%%%%%%%%%%%%%%%%%%%%%%%%%%%%%%%%%%%%%%%%%%%%%%%%%%%%%%%%%%%%%
\begin{align}
\Delta\Omega^{T}_{\rm grav}&\equiv \Omega^{T}_{\rm grav}-\Omega_{\rm ph}=-2\alpha_{2}\left(\frac{d-1}{d-4}\right)\Omega_{\textrm{ph}(1)}~,
\\
\Delta\Omega^{V}_{\rm grav}&\equiv\Omega^{V}_{\rm grav}-\Omega_{\rm ph}=(d-1)\alpha_2 \Omega_{\textrm{ph}(1)}~,
\\
\Delta\Omega^{S}_{\rm grav}&\equiv\Omega^{S}_{\rm grav}-\Omega_{\rm ph}=2\alpha_{2}(d-1)\Omega_{\textrm{ph}(1)}~, 
\end{align}
%%%%%%%%%%%%%%%%%%%%%%%%%%%%%%%%%%%%%%%%%%%%%%%%%%%%%%%%%%%%%%
%$\Omega_c^A$  can also be expanded in terms of the coupling constant as,
%\begin{equation}
%\Omega_c^A= \Omega_{c0}^A+\Omega_{c1}^A \lambda_{GB}+\mathcal{O}(\lambda_{GB}^2) ,
%\end{equation}
%and
% \begin{eqnarray}
% &\Omega_{c0}^A=\sqrt{\frac{d-3}{d-1}} \left(\frac{2}{(d-1) \mu }\right)^{\frac{1}{d-3}} ,\\
% &\Omega_{c1}^T= \frac{3 (d-2)}{d-4}\Omega_{c1}^0\\
% &\Omega_{c1}^V= (2 - d)~\Omega_{c1}^0\\
% &\Omega_{c1}^V= (3 - 2d)~\Omega_{c1}^0,
% %&\Omega_{c1}^A= \zeta^A 3 (d-2) 2^{\frac{3}{d-3}} \sqrt{\frac{d-3}{d-1}}   \mu  \left(\frac{1}{(d-1) \mu }\right)^{\frac{d}{d-3}} ,
% \end{eqnarray}
where, $\Omega_{\textrm{ph}(1)}$ is the correction to the angular velocity of circular photon geodesics at the first order in the Gauss-Bonnet coupling $\alpha_{2}$, which reads,
%%%%%%%%%%%%%%%%%%%%%%%%%%%%%%%%%%%%%%%%%%%%%%%%%%%%%%%%%%%%%%
\begin{equation}
\Omega_{\textrm{ph}(1)}=\frac{2^{\frac{d}{d-3}} (d-4) \sqrt{d-3}}{(d-1)^{3/2}\left[(d-1) \mu \right]^{\frac{3}{d-3}}}~.
\end{equation}
%%%%%%%%%%%%%%%%%%%%%%%%%%%%%%%%%%%%%%%%%%%%%%%%%%%%%%%%%%%%%%
%$\zeta^T=1$, $\zeta^V=-(d-4)/3$, $\zeta^S=-\frac{(d-4)(2d-3)}{3 (d-2)}$ and $\zeta^0=\frac{(d-4)}{3 (d-2)}$.
Note that in $d=4$ this term identically vanishes, since the Gauss-Bonnet term is absent in four dimensions. 

Since the circular null geodesics of both the physical and effective metric correspond to the peak of the effective potential, these are unstable geodesics of the corresponding metric. The timescale of this instability associated with these orbits is analyzed in terms of the Lyapunov exponents, which measure the rate at which nearby trajectories converge or diverge. A positive Lyapunov exponent implies divergence of nearby geodesics. 
%One generally starts the analysis by writing the equation of motion in a  schematic form,
%\begin{equation}
% \frac{dX_i}{dt}=H_i(X_j),
% \end{equation}
% which, when linearized about a certain orbit, takes the form,
% \begin{equation}
% \frac{d\delta X_i}{dt}=K_{ij}\delta X_j(t),
% \end{equation}
% where
% \begin{equation}
% K_{ij}(t)=\frac{\partial H_i}{\partial X_j}\Biggr |_{X_i(t)} ,
% \end{equation}
% is the infinitesimal evolution matrix with $\delta X_i$ denoting the infinitesimal orbital deviation. The solution to the linearized equation can be expressed as,
% \begin{equation}
% \delta X_i(t)=L_{ij}(t) \delta X_j(0) ,
% \end{equation}
% in terms of the evolution matrix $L_{ij}$, which must obey
% \begin{equation}
% \dot{L}_{ij}(t)=K_{im}L_{mj}(t), \qquad L_{ij}(0)=\delta_{ij}~.
% \end{equation}
% The principal Lyapunov exponent is then given by,
% \begin{equation}
% \lambda= \lim_{t\rightarrow \infty} \frac{1}{t} \log\left(\frac{L_{jj}(t)}{L_{jj}(0)}\right)~.
% \end{equation}
The principal Lyapunov exponent in the case of unstable circular null geodesics, namely the gravitonsphere, is given by~\cite{cardoso2009PRD},
%%%%%%%%%%%%%%%%%%%%%%%%%%%%%%%%%%%%%%%%%%%%%%%%%%%%%%%%%%%%%%
\begin{equation}\label{eq:lyapunov}
\lambda_{\rm grav}^{A}=\left(-\frac{f(r)^2 \partial_r^2V_{\text{eff}}^A (r)}{2 V_{\text{eff}}^{A}(r)} \right)^{1/2}_{r=r_{\rm grav}^{A}}~.
\end{equation}
%%%%%%%%%%%%%%%%%%%%%%%%%%%%%%%%%%%%%%%%%%%%%%%%%%%%%%%%%%%%%%
Expanding the Lyapunov exponent in powers of the Gauss-Bonnet coupling constant, we note that it differs from the Lyapunov exponent at the photon sphere, only in the second order in the coupling constant, such that for tensor, vector and scalar modes, we have 
%%%%%%%%%%%%%%%%%%%%%%%%%%%%%%%%%%%%%%%%%%%%%%%%%%%%%%%%%%%%%%
\begin{align}\label{eq:lyapunov-ser}
\Delta\lambda^{T}_{\rm grav}&\equiv\lambda^{T}_{\rm grav}-\lambda_{\rm ph}
\nonumber
\\
&=-\alpha_{2}^{2}\frac{4(d-1)^{2}\left(d^3-8 d^2+16 d+9\right)}{(d-4)^2\left(3 d^2-24 d+28\right)}\lambda_{\textrm{ph}(2)}~,
\\
\Delta\lambda^{V}_{\rm grav}&\equiv\lambda^{V}_{\rm grav}-\lambda_{\rm ph}=\alpha_{2}^{2}\frac{(d-13) (d-1)^2}{3 d^2-24 d+28}\lambda_{\textrm{ph}(2)}~,
\\
\Delta\lambda^{S}_{\rm grav}&\equiv\lambda^{S}_{\rm grav}-\lambda_{\rm ph}
\nonumber
\\
&=-\alpha_{2}^{2}\frac{12(d-1)^{2}(3d-5)}{(d-2)\left(3 d^2-24 d+28\right)}\lambda_{\textrm{ph}(2)}~,
\end{align}
%%%%%%%%%%%%%%%%%%%%%%%%%%%%%%%%%%%%%%%%%%%%%%%%%%%%%%%%%%%%%%
where, $\lambda_{\textrm{ph}(2)}$ is the correction to the Lypanunov exponent of the photon sphere at the second order in the Gauss-Bonnet coupling, which reads,
%%%%%%%%%%%%%%%%%%%%%%%%%%%%%%%%%%%%%%%%%%%%%%%%%%%%%%%%%%%%%%
\begin{equation}
\lambda_{\textrm{ph}(2)}=\frac{2^{\frac{8-d}{d-3}} (d-4)^2 (d-3) \left[3 (d-8) d+28\right]}{(d-1)^{\frac{5(d-1)}{2(d-3)}}\mu^{\frac{5}{(d-3)}}}~.
\end{equation}
%%%%%%%%%%%%%%%%%%%%%%%%%%%%%%%%%%%%%%%%%%%%%%%%%%%%%%%%%%%%%%
% we get, 
% \begin{eqnarray}
% \lambda_{c0}^A&=&\frac{ (d-3) \left(\frac{2}{(d-1) \mu }\right)^{\frac{1}{d-3}}}{\sqrt{d-1}}\\
% %
% \lambda_{c1}^A&=&\frac{(d-4) (d-3) (d-2) \lambda  \mu  \left(\frac{2}{(d-1) \mu }\right)^{\frac{d}{d-3}}}{\sqrt{d-1}}
% \end{eqnarray}
% %\begin{widetext}
% \begin{eqnarray}
% \lambda_{c2}^T&=& 2^{\frac{d+2}{d-3}}\frac{ (d-3) (d (d ((d-4) d (4 d-21)+144)-616)+484)}{(d-1)^{\frac{5(d-1)}{2(d-3)}}\mu^{5/(d-3)}},\\
% %
% \lambda_{c2}^V&=&-\frac{2^{\frac{d+2}{d-3}} (d-4)^2 (d-3) (d ((d-18) d+51)-41) }{(d-1)^{\frac{5(d-1)}{2(d-3)}}\mu^{5/(d-3)}},\\
% %
% \lambda_{c2}^S&=&\frac{2^{\frac{d+2}{d-3}} (d-4)^2 (d-3) (d (3 d (13 d-54)+232)-116)}{(d-2) (d-1)^{\frac{5(d-1)}{2(d-3)}}\mu^{5/(d-3)}}~,\\
% %
% \lambda_{c2}^0&=&\frac{2^{\frac{d+2}{d-3}} (d-4)^2 (d-3) (3 (d-8) d+28)}{(d-1)^{\frac{5(d-1)}{2(d-3)}}\mu^{5/(d-3)}} .
% \end{eqnarray}
% %\end{widetext}
The expression for dominant terms in the expansion of the angular velocity and the Lyapunov exponent of the unstable null cicular geodesic of the physical metric are given in Appendix~\ref{AppA}.

%%%%%%%%%%%%%%%%%%%%%%%%%%%%%%%%%%%%%%%%%%%%%%%%%%%%%%%%%%%%%%%%%%%%
%%%%%%%%%%%%%%%%%%%%%%%%%%%%%%%%%%%%%%%%%%%%%%%%%%%%%%%%%%%%%%%%%%%%
%%%%%%%%%%%%%%%%%%%%%%%%%%%%%%%%%%%%%%%%%%%%%%%%%%%%%%%%%%%%%%%%%%%%
\subsubsection{Pure Lovelock gravity}\label{sec:2Bii}

In pure Lovelock gravity, it follows that the functions $c_A(r)$ do not have any radial dependence, rather they are constants for all choices of $A$ (see, Eq.\eqref{eq:cA-PLL-1}, Eq.\eqref{eq:cA-PLL-2}, and Eq.\eqref{eq:cA-PLL-3} for details), and hence the location of the photon sphere coincides with the location of the graviton-sphere (for all the scalar, vector and tensor types), yielding,
%%%%%%%%%%%%%%%%%%%%%%%%%%%%%%%%%%%%%%%%%%%%%%%%%%%%%%%%%%%%%%
\begin{equation}
r_{\rm grav}^A=r_{+}\left(\frac{d-1}{2 N}\right)^{\frac{N}{d-2 N-1}}~.
\end{equation}
%%%%%%%%%%%%%%%%%%%%%%%%%%%%%%%%%%%%%%%%%%%%%%%%%%%%%%%%%%%%%%
Also, there is only one maxima to the effective potential, and hence, there exists no instability for pure Lovelock black holes. Furthermore, from Eq.\eqref{eq:lyapunov}, we see that the principal Lyapunov exponent for the scalar, vector, and tensor perturbations matches with that of the photon sphere, which in $d$ dimensions for $N$th order pure Lovelock black hole spacetime reads,
%%%%%%%%%%%%%%%%%%%%%%%%%%%%%%%%%%%%%%%%%%%%%%%%%%%%%%%%%%%%%%
\begin{equation}
    %\lambda_c^A=\frac{ 2^{\frac{N}{d-2 N-1}}(d-1-2N) \left(\frac{d-1}{N}\right)^{-\frac{d-1}{2(d-2 N-1)}}}{r_s N} .
    \lambda_c^A=\frac{d-1-2N}{\sqrt{ N (d-1)}  r_c^A}
\end{equation}
%%%%%%%%%%%%%%%%%%%%%%%%%%%%%%%%%%%%%%%%%%%%%%%%%%%%%%%%%%%%%%
On the other hand, the angular velocity of the null circular geodesic of the effective metric is different from that of the physical metric, such that,
%%%%%%%%%%%%%%%%%%%%%%%%%%%%%%%%%%%%%%%%%%%%%%%%%%%%%%%%%%%%%%
\begin{align}
\Omega_{\rm grav}^{A}&=\sqrt{c_A}\Omega_{\rm ph}
\nonumber
\\
&=\frac{\sqrt{c_A}}{r_{+}}\left(\frac{2N}{d-1}\right)^{\frac{N}{d-2N-1}}\sqrt{1-\frac{2N}{d-1}}~,
\end{align}
%%%%%%%%%%%%%%%%%%%%%%%%%%%%%%%%%%%%%%%%%%%%%%%%%%%%%%%%%%%%%%
where $A\neq 0$ and depicts the angular velocity of scalar, vector and tensor spheres. As evident this is different from the corresponding expression for the photon sphere. Thus, as far as pure Lovelock gravity is concerned, neither the location of the graviton sphere nor the instability timescale of the graviton sphere are different from that of the photon sphere. This is in stark contrast with the general Lovelock case, e.g., in the case of Einstein-Gauss-Bonnet gravity considered above.

%%%%%%%%%%%%%%%%%%%%%%%%%%%%%%%%%%%%%%%%%%%%%%%%%%%%%%%%%%%%%%%%%%%%%%%%%%%%%%%%%%%
%%%%%%%%%%%%%%%%%%%%%%%%%%%%%%%%%%%%%%%%%%%%%%%%%%%%%%%%%%%%%%%%%%%%%%%%%%%%%%%%%%%
%%%%%%%%%%%%%%%%%%%%%%%%%%%%%%%%%%%%%%%%%%%%%%%%%%%%%%%%%%%%%%%%%%%%%%%%%%%%%%%%%%%
%%%%%%%%%%%%% Section 3 %%%%%%%%%%%%%%%%%%
%%%%%%%%%%%%%%%%%%%%%%%%%%%%%%%%%%%%%%%%%%%%%%%%%%%%%%%%%%%%%%%%%%%%%%%%%%%%%%%%%%%
%%%%%%%%%%%%%%%%%%%%%%%%%%%%%%%%%%%%%%%%%%%%%%%%%%%%%%%%%%%%%%%%%%%%%%%%%%%%%%%%%%%
%%%%%%%%%%%%%%%%%%%%%%%%%%%%%%%%%%%%%%%%%%%%%%%%%%%%%%%%%%%%%%%%%%%%%%%%%%%%%%%%%%%
\section{Gravitonspheres and the QNMs for gravitational perturbations}\label{sec:3}

Having discussed the location and the properties of the gravitonsphere, let us concentrate on the QNMs associated with gravitational perturbation and their connection to the gravitonsphere. Following~\cite{kodama2003PTP}, linearized gravitational perturbation in Lovelock theories (in general, linearized perturbation of a symmetric rank-2 tensor) can be decomposed in scalar, vector and tensor components based on how they transform under parity. The tensor part of the perturbation arises solely from higher dimensions, while the scalar part is parity even and the vector part is parity odd, and all of these perturbations decouple at linear order. Given the static and spherically symmetric background, the master function $\Psi_{A}$ for the scalar, vector and tensor perturbations can be decomposed as,
%%%%%%%%%%%%%%%%%%%%%%%%%%%%%%%%%%%%%%%%%%%%%%%%%%%%%%%%%%%%%%
\begin{equation}
\Psi_{A}=\sum_{\ell m_{1}\cdots m_{d-3}}\int dt e^{-i\omega t}R_{A}(r)Y_{\ell m_{1}\cdots m_{d-3}}(\theta, \phi_{1},\cdots \phi_{d-3})~,
\end{equation}
%%%%%%%%%%%%%%%%%%%%%%%%%%%%%%%%%%%%%%%%%%%%%%%%%%%%%%%%%%%%%%
where, $Y_{\ell m_{1}\cdots m_{d-3}}(\theta, \phi_{1},\cdots \phi_{d-3})$ are the harmonics over the $(d-2)$ sphere. The radial function $R_{A}(r)$ satisfies the following second order differential equations for the tensor, vector and scalar modes, in terms of the tortoise coordinate as~\cite{takahashi2010PTP}, 
%%%%%%%%%%%%%%%%%%%%%%%%%%%%%%%%%%%%%%%%%%%%%%%%%%%%%%%%%%%%%%
\begin{equation}\label{eq:schrodinger}
\frac{d^{2}R_{A}}{dr_{*}^{2}}+\left(\omega^{2}-V_{A}\right)R_{A}=0~.
\end{equation}
%%%%%%%%%%%%%%%%%%%%%%%%%%%%%%%%%%%%%%%%%%%%%%%%%%%%%%%%%%%%%%
Note that the tortoise coordinate $r_{*}$ is defined as $(dr/dr_{*})=f(r)$, and maps the the semi-infinite region $[r_{+},\infty)$, to $(-\infty,\infty)$. Therefore, the QNM frequencies $\omega$ are determined by purely ingoing boundary condition at the horizon and purely outgoing boundary condition at infinity, along with the potential $V_{A}(r)$, which in the eikonal limit, for all three types of gravitational perturbations can be approximated as~\cite{takahashi2010PTP2, takahashi2010PTP},
%%%%%%%%%%%%%%%%%%%%%%%%%%%%%%%%%%%%%%%%%%%%%%%%%%%%%%%%%%%%%%
\begin{equation}\label{eq:Veff-per}
V_{A}(r)=\ell^{2}\left[\frac{f_{A}(r)}{r^2}+\mathscr{O}\left(\frac{1}{\ell}\right)\right]~.
\end{equation}
%%%%%%%%%%%%%%%%%%%%%%%%%%%%%%%%%%%%%%%%%%%%%%%%%%%%%%%%%%%%%%
The function $f_{A}(r)$, for scalar, vector and tensor perturbations take the following forms,
%%%%%%%%%%%%%%%%%%%%%%%%%%%%%%%%%%%%%%%%%%%%%%%%%%%%%%%%%%%%%%
\begin{align}\label{eq:f_A}
f_{S}(r)&=\frac{f(r)r\left[2{T'(r)}^{2}-T(r)T''(r)\right]}{(d-2)T'(r)T(r)}~,
\\
f_{V}(r)&=\frac{f(r)rT'(r)}{(d-3)T(r)}~; 
\quad 
f_{T}(r)=\frac{f(r)rT''(r)}{(d-4)T'(r)}~, 
\end{align}
%%%%%%%%%%%%%%%%%%%%%%%%%%%%%%%%%%%%%%%%%%%%%%%%%%%%%%%%%%%%%%
where `prime' denotes the derivative with respect to the radial coordinate $r$, and the function $T(r)$ reads as,
%%%%%%%%%%%%%%%%%%%%%%%%%%%%%%%%%%%%%%%%%%%%%%%%%%%%%%%%%%%%%%
\begin{equation}
T(r)\equiv r^{d-3}\frac{dW[\psi]}{d\psi}~,
\end{equation}
%%%%%%%%%%%%%%%%%%%%%%%%%%%%%%%%%%%%%%%%%%%%%%%%%%%%%%%%%%%%%%
with $W[\psi(r)]$ defined in Eq.~\eqref{eq:W}. Using the above expressions for $f_A(r)$ and $T(r)$ along with $W[\psi]$ from Eq.~\eqref{eq:W} and its radial derivative,
%%%%%%%%%%%%%%%%%%%%%%%%%%%%%%%%%%%%%%%%%%%%%%%%%%%%%%%%%%%%%%
\begin{equation}
\frac{dW[\psi]}{dr}=-(d-1)\frac{W[\psi]}{r\left(\frac{dW}{d\psi}\right)}~,   
\end{equation}
%%%%%%%%%%%%%%%%%%%%%%%%%%%%%%%%%%%%%%%%%%%%%%%%%%%%%%%%%%%%%%
it is straightforward to verify that the functions $f_{A}(r)$ and the metric functions $c_{A}(r)$ given in Eq.~\eqref{eq:cA}, are related by,
%%%%%%%%%%%%%%%%%%%%%%%%%%%%%%%%%%%%%%%%%%%%%%%%%%%%%%%%%%%%%%
\begin{equation}\label{eq:eqv-f}
f_A(r)=f(r) c_A(r)~,
\end{equation}
%%%%%%%%%%%%%%%%%%%%%%%%%%%%%%%%%%%%%%%%%%%%%%%%%%%%%%%%%%%%%%
for both the full Lovelock gravity as well as for the pure Lovelock theories. Thus it immediately follows that in the eikonal limit, $V_{A}\approx V^{A}_{\rm eff}$ and hence maxima of the potential governing the perturbation 

In order to calculate the QNMs in the eikonal limit, we resort to the WKB method developed in \cite{mashhoon1982MG,schutz1985APJL,iyer1986PRD,iyer1986PRD1}, which yields 
%%%%%%%%%%%%%%%%%%%%%%%%%%%%%%%%%%%%%%%%%%%%%%%%%%%%%%%%%%%%%%
\begin{equation}\label{eq:WKB}
\frac{\omega^{2}-V_{A}(r_{\rm grav}^{A})}{\sqrt{-2(d^{2}V_{A}/dr_{*}^{2})_{\rm grav}}}=i\left(n+\frac{1}{2}\right)~.
\end{equation}
%%%%%%%%%%%%%%%%%%%%%%%%%%%%%%%%%%%%%%%%%%%%%%%%%%%%%%%%%%%%%%
All the quantities appearing on the left hand side of the above expression are to be evaluated at the maxima of the potential $V_{A}$. This coincides with the location of the gravitonspheres. This is because, in the eikonal limit, it follows that $V_{A}(r)\approx V_{\text{eff}}^{A}$, and hence the maxima of $V_{\text{eff}}^{A}$ coincides with the maxima of $V_{A}$, yielding the location of unstable null geodesics in the effective spacetime. 

For small values of the coupling parameters, in Lovelock gravity, the perturbing potential $V_A$ for all three types of gravitational perturbation has a single maximum, as depicted in Fig. \ref{fig:potential} (see also~\cite{reall2014CQG}), and hence the WKB formula works well. However, for larger coupling parameters, the potential governing scalar part of gravitational perturbation develops a minimum, signalling existence of stable gravitonspheres and hence instabilities. The WKB formula, thus, is not applicable for scalar mode of gravitational perturbation in the large coupling parameter regime.  

Given Eq.~\eqref{eq:WKB}, one can determine the real and the imaginary part of the QNM frequencies. It follows that the angular velocity of the unstable null geodesics in the effective metric matches with the real part of the QNM frequencies, and the Lyapunov exponent associated with unstable null geodesics of the effective metric is given by the imaginary part of the QNM frequencies. When expressed explicitly, we obtain,
%%%%%%%%%%%%%%%%%%%%%%%%%%%%%%%%%%%%%%%%%%%%%%%%%%%%%%%%%%%%%%
\begin{equation}\label{eq:wkb-final}
\omega^{A}_{\text{QNM}}=\ell \Omega_{\rm grav}^{A}-i\left(n+\frac{1}{2}\right)\Bigr|\lambda_{\rm grav}^{A}\Bigl|~.
\end{equation}
%%%%%%%%%%%%%%%%%%%%%%%%%%%%%%%%%%%%%%%%%%%%%%%%%%%%%%%%%%%%%%
for the QNM frequencies of the gravitational perturbation in the eikonal limit. 

Similarly, for pure Lovelock gravity, the relations presented in Eqs.~(\ref{eq:eqv-f},\ref{eq:WKB},\ref{eq:wkb-final}) holds in the eikonal limit. Since, in this case, the Lyapunov exponent at the unstable circular null geodesics of the physical and the effective metric are the same and only the angular velocity of the null geodesics differ, it follows that the scalar, vector and tensor QNMs have the same damping rate but different oscillating frequencies.

%%%%%%%%%%%%%%%%%%%%%%%%%%%%%%%%%%%%%%%%%%%%%%%%%%%%%%%%%%%%%%%%%%%%%%%%%%%%%%%%%%%
%%%%%%%%%%%%%%%%%%%%%%%%%%%%%%%%%%%%%%%%%%%%%%%%%%%%%%%%%%%%%%%%%%%%%%%%%%%%%%%%%%%
%%%%%%%%%%%%%%%%%%%%%%%%%%%%%%%%%%%%%%%%%%%%%%%%%%%%%%%%%%%%%%%%%%%%%%%%%%%%%%%%%%%
\section{Cross-Validation: Comparison of Analytical and Numerical Results}\label{sec:4}

%%%%%%%%%%%%%%%%%%%%
%%%%%%%%%%%%%%%%%%%%
%%%%%%%%%%%%%%%%%%%%
%%%%%%%%%%%%%%%%%%%%
\begin{figure*}[!tb]
\centering
\includegraphics[width=0.32\textwidth]{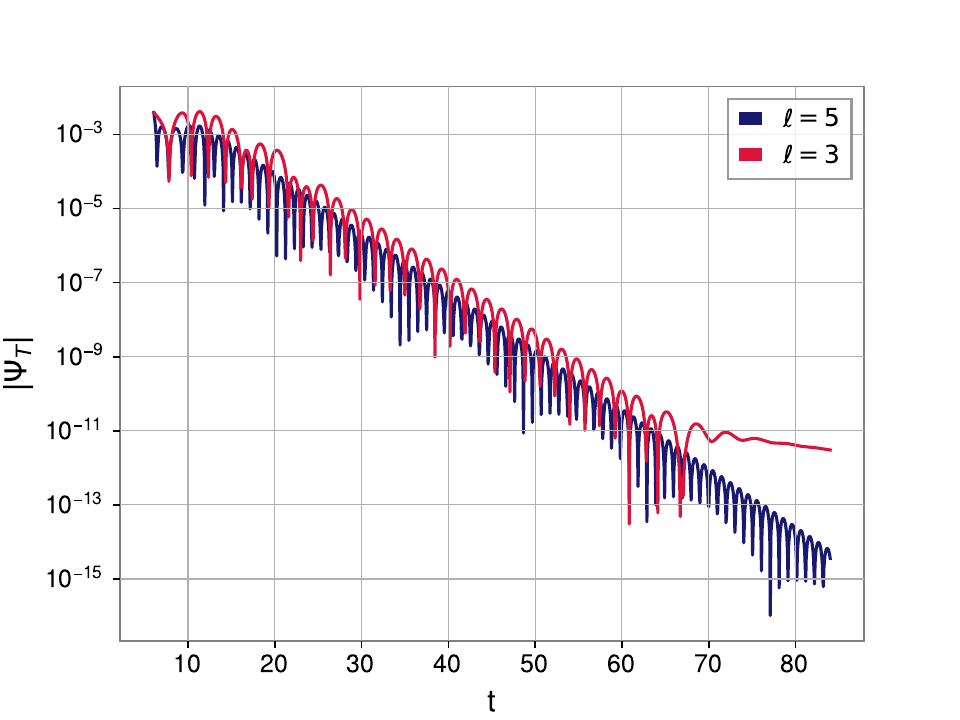}
\includegraphics[width=0.32\textwidth]{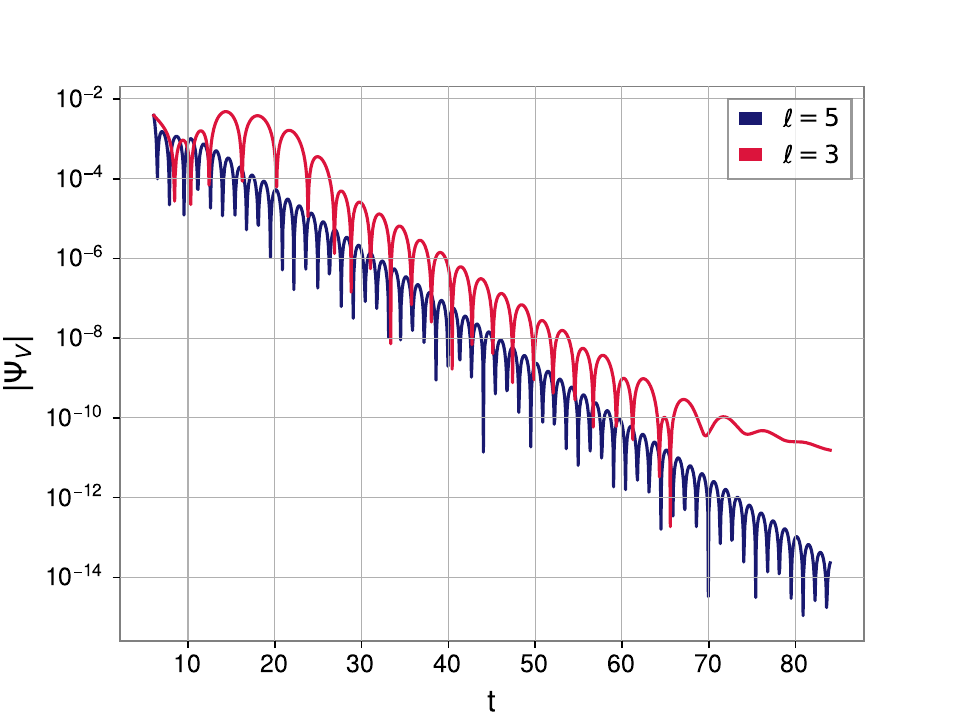}
\includegraphics[width=0.32\textwidth]{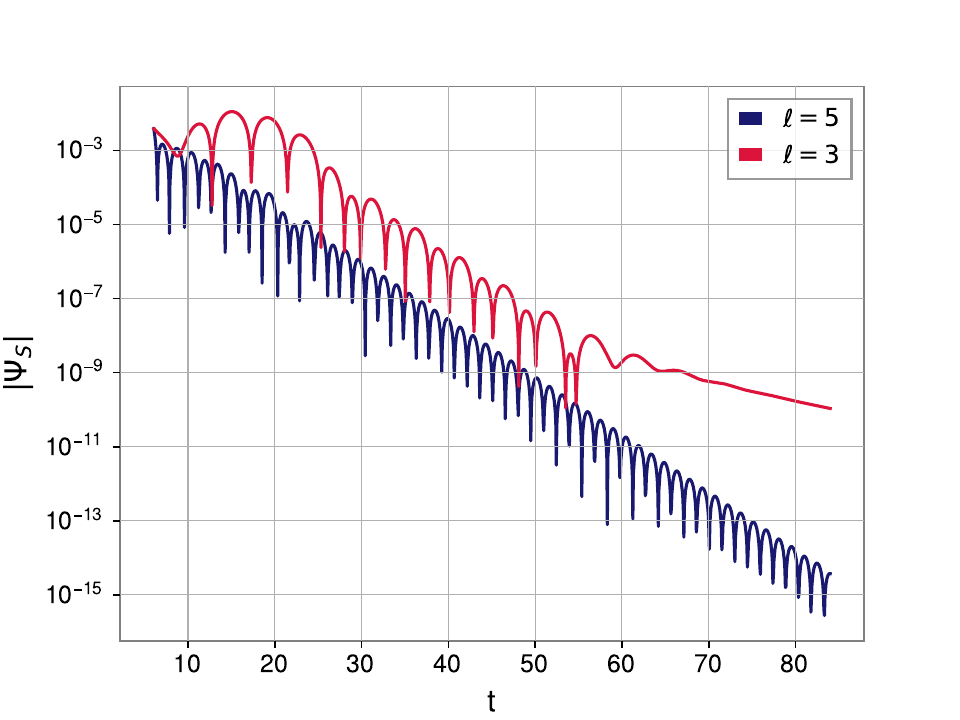}
\caption{Time evolution of tensor, vector and scalar modes of gravitational perturbations have been presented for $\ell=3,5$ and for the choice of Gauss-Bonnet coupling parameter $\alpha_2=0.1$ in five dimensions.}
\label{fig:time_domain_waveform}
\end{figure*}
%%%%%%%%%%%%%%%%%%%%
%%%%%%%%%%%%%%%%%%%%
%%%%%%%%%%%%%%%%%%%%
%%%%%%%%%%%%%%%%%%%%

In the previous sections, we have established that the QNM frequencies in the eikonal limit are closely related to the properties of unstable circular null geodesics of the effective graviton metric in Lovelock theories of gravity. In this section, we further strengthen this analogy by comparing the numerically computed QNMs with the analytical expression given by Eq.~\eqref{eq:wkb-final} in the context of BHs in five-dimensional Einstein-Gauss-Bonnet gravity and pure Lovelock BHs. 

For the computation of the QNMs, we use the time-domain evolution method, where we integrate the master wave equation in the $(u=t-r_*,\,v=t+r_*)$ coordinate system. In this light-cone coordinates, the radial perturbation equation, presented in Eq.~\eqref{eq:schrodinger}, albeit in the time domain, can be written as, 
%%%%%%%%%%%%%%%%%%%%%%%%%%%%%%%%%%%%%%%%%%%%%%%%%%%%%%%%%%%%%%
\begin{equation}\label{eq:uv-eqn}
\left(4\frac{\partial^2}{\partial u \partial v} + V_A(u,v) \right)\Psi_A(u,v)=0
\end{equation}
%%%%%%%%%%%%%%%%%%%%%%%%%%%%%%%%%%%%%%%%%%%%%%%%%%%%%%%%%%%%%%
The above equation can be numerically solved in a straightforward manner on a null grid by adapting an appropriate discretization scheme. In particular, one relates the perturbation variable at a neighbouring point $\Psi_{A}(u_{0}+h,v_{0}+h)$ with $\Psi_{A}(u_{0},v_{0})$:
%%%%%%%%%%%%%%%%%%%%%%%%%%%%%%%%%%%%%%%%%%%%%%%%%%%%%%%%%%%%%%
\begin{align}\label{discrit}
&\Psi_{A}(u_0+h,\, v_0+h) = \Psi_{A}(u_0+h, v_0)+\Psi_{A}(u_0, v_0+h)  
\nonumber
\\
&\qquad -\Psi_{A}(u_0,v_0)-\frac{h^2}{8}\Big[V_{A}(u_0+h,v_0)\Psi_{A}(u_0+h,v_0) 
\nonumber
\\
&\qquad + V_{A}(u_0,v_0+h)\Psi_{A}(u_0,v_0+h)\Big]~.
\end{align}
%%%%%%%%%%%%%%%%%%%%%%%%%%%%%%%%%%%%%%%%%%%%%%%%%%%%%%%%%%%%%%
The above relation allows us to obtain the value of the field $\Psi_{A}$ on the entire null grid by starting from an initial data at a specific point $(u_{0},v_{0})$. For the perturbation variable, $\Psi_{A}$, we take its initial choice to be a Gaussian: $\Psi_{A}(u_{0},v_0)=\exp\left[-(v_{0}-10)^{2}/18\right]$. Therefore, the subsequent evolution of this wave packet throughout the null plane provides the time domain waveform, as the above analysis provides the value of the field as a function of time i.e., $\Psi_{A}(t_0), \Psi_{A}(t_0+h), \Psi_{A}(t_0+2h), \cdots$ etc., which we use to compute the QNMs by a Prony fit algorithm~\cite{Berti:2007dg, konoplya:2011qq}.

The result of the above analysis has been presented in Fig. \ref{fig:time_domain_waveform}. Here we show the time domain profile of the tensor, vector and scalar part of the gravitational perturbation for five-dimensional Einstein-Gauss-Bonnet gravity associated with $\ell=3$ and $\ell=5$ modes. Using the above time-domain profile, we employ the Prony fit method to compute the QNM frequencies for various modes. In Table~\ref{tensor_perturbation_qnm}, Table~\ref{vector_perturbation_qnm} and Table~\ref{scalar_perturbation_qnm}, we present a comparison between the QNM frequencies obtained through eikonal approximation and by the Prony fit algorithm for tensor, vector and scalar modes of gravitational perturbation. This comparison clearly illustrates the matching of the eikonal QNM frequencies obtained analytically with the numerical results and hence confirms our proposed relationship between the eikonal QNMs and the properties of the graviton sphere. It turns out that an identical result holds for pure Lovelock theories of gravity as well, for which we have illustrated the above relationship in the context of pure Gauss-Bonnet gravity in $d=7,8$ for scalar part of gravitational perturbation in Table~\ref{pure_lovelock_qnm}.
  
%%%%%%%%%%%%%%%%%%%%
%%%%%%%%%%%%%%%%%%%%
%%%%%%%%%%%%%%%%%%%%
%%%%%%%%%%%%%%%%%%%%
\begin{table}[H]
\begin{centering}
\begin{tabular}{|c|c|c|c|c|}
\hline 
$\ell$  &  $\alpha_{2}$ & Eikonal  & Numerical  \tabularnewline
\hline 
\hline 
\multirow{4}{*}{$2$}  & \multirow{1}{*}{$0.05$}  & $1.123956 - 0.359332 i$  & $1.03319 - 0.347046 i$ \tabularnewline

\cline{2-5} 
 &  \multirow{1}{*}{$0.1$} & $1.251348 - 0.365687 i
$  & $1.16771 - 0.348191 i$  \tabularnewline

\cline{2-5} \cline{2-5}  
  & \multirow{1}{*}{$0.15$} & $1.370181 - 0.356663 i$  & $1.30135 - 0.352978 i$  \tabularnewline
 
\cline{2-5}
 &  \multirow{1}{*}{$0.2$}  & $1.481061 - 0.337945 i$   & $1.42085 - 0.336404 i$  \tabularnewline
 
\cline{5-5} 
\hline 

\cline{5-5} 
\hline 
\multirow{4}{*}{$3$}  & \multirow{1}{*}{$0.05$}  & $1.68593 - 0.35933 i$ & $1.62406 - 0.354185 i$ \tabularnewline

\cline{2-5} 
 &  \multirow{1}{*}{$0.1$} & $1.87702 - 0.36569 i$  & $1.81868 - 0.359658 i$ \tabularnewline

\cline{2-5} \cline{3-5}  
  & \multirow{1}{*}{$0.15$} & $2.05527 - 0.35666 i$  & $2.00542 - 0.355071 i$ \tabularnewline
 
\cline{2-5}
 &  \multirow{1}{*}{$0.2$}  & $2.22159 - 0.33794 i$  & $2.18249 - 0.339964 i$\tabularnewline
 
\cline{5-5} 
\hline 

\cline{5-5} 
\hline 
\multirow{4}{*}{$5$}  & \multirow{1}{*}{$0.05$}  & $2.80989 - 0.35933 i$  & $2.77309 - 0.359181 i$\tabularnewline

\cline{2-5} 
 &  \multirow{1}{*}{$0.1$} & $3.12837 - 0.36569 i$  & $3.09315 - 0.366243 i$\tabularnewline

\cline{2-5} \cline{3-5}  
  & \multirow{1}{*}{$0.15$} & $3.42544 - 0.35666 i$ & $3.39689 - 0.359251 i$\tabularnewline
 
\cline{2-5}
 &  \multirow{1}{*}{$0.2$}  & $3.70265 - 0.33794 i$ & $3.68138 - 0.34138 i$ \tabularnewline
\cline{5-5} 
\hline 

\cline{5-5} 
\hline 
\multirow{4}{*}{$10$}  & \multirow{1}{*}{$0.05$}  & $5.61978 - 0.35933 i$  & $5.60721 - 0.36148 i$\tabularnewline

\cline{2-5} 
 &  \multirow{1}{*}{$0.1$} & $6.25674 - 0.36569 i$  & $6.24759 - 0.368088 i$\tabularnewline

\cline{2-5} \cline{3-5}  
  & \multirow{1}{*}{$0.15$} & $6.85088 - 0.35666 i$ & $6.84814 - 0.35917 i$\tabularnewline
 
\cline{2-5}
 &  \multirow{1}{*}{$0.2$}  & $7.40530 - 0.33794 i$ & $7.40928 - 0.34013 i$ \tabularnewline
\cline{5-5} 
\hline 

\end{tabular}
\par\end{centering}
\caption{In this table we compare the fundamental QNM frequncies ($n = 0$) for different $\ell$ and $\alpha_{2}$ of the tensor mode associated with gravitational perturbation of Einstein-Gauss-Bonnet BH in $d=5$ dimension using eikonal as well as numerical method.}
\label{tensor_perturbation_qnm}
\end{table}
%%%%%%%%%%%%%%%%%%%%
%%%%%%%%%%%%%%%%%%%%
%%%%%%%%%%%%%%%%%%%%
%%%%%%%%%%%%%%%%%%%%

%%%%%%%%%%%%%%%%%%%%
%%%%%%%%%%%%%%%%%%%%
%%%%%%%%%%%%%%%%%%%%
%%%%%%%%%%%%%%%%%%%%
\begin{table}[!tb]
\begin{centering}
\begin{tabular}{|c|c|c|c|c|}
\hline 
$\ell$  &  $\alpha_{2}$ & Eikonal  & Numerical  \tabularnewline
\hline 
\hline 
\multirow{4}{*}{$2$}  & \multirow{1}{*}{$0.05$}  & $0.965820 - 0.342839 i
$  & $0.878499 - 0.325282 i$ \tabularnewline

\cline{2-5} 
 &  \multirow{1}{*}{$0.1$} & $0.937110 - 0.335635 i
$  & $0.844024 - 0.312625 i$  \tabularnewline

\cline{2-5} \cline{2-5}  
  & \multirow{1}{*}{$0.15$} & $0.912530 - 0.330419 i$  & $0.808529 - 0.292533 i$  \tabularnewline
 
\cline{2-5}
 &  \multirow{1}{*}{$0.2$}  & $0.891143 - 0.326368 i$   & $0.798821 - 0.299321 i$  \tabularnewline
 
\cline{5-5} 
\hline 

\cline{5-5} 
\hline 
\multirow{4}{*}{$3$}  & \multirow{1}{*}{$0.05$}  & $1.44873 - 0.34284 i$ & $1.384533 - 0.335453 i$ \tabularnewline

\cline{2-5} 
 &  \multirow{1}{*}{$0.1$} & $1.40567 - 0.33564 i$  & $1.340331 - 0.325587 i$ \tabularnewline

\cline{2-5} \cline{3-5}  
  & \multirow{1}{*}{$0.15$} & $1.368795 - 0.330419 i$  & $1.302042 - 0.318717 i$ \tabularnewline
 
\cline{2-5}
 &  \multirow{1}{*}{$0.2$}  & $1.336714 - 0.326368 i$  & $1.268881 - 0.313745 i$\tabularnewline
 
\cline{5-5} 
\hline 

\cline{5-5} 
\hline 
\multirow{4}{*}{$5$}  & \multirow{1}{*}{$0.05$}  & $2.41455 - 0.34284 i$  & $2.375833 - 0.342007 i$\tabularnewline

\cline{2-5} 
 &  \multirow{1}{*}{$0.1$} & $2.34278 - 0.33564 i$  & $2.302414 - 0.333438 i$\tabularnewline

\cline{2-5} \cline{3-5}  
  & \multirow{1}{*}{$0.15$} & $2.28133 - 0.33042 i$ & $2.240277 - 0.327583 i$\tabularnewline
 
\cline{2-5}
 &  \multirow{1}{*}{$0.2$}  & $2.22786 - 0.32637 i$ & $2.186331 - 0.323122 i
$ \tabularnewline
\cline{5-5} 
\hline 

\cline{5-5} 
\hline 
\multirow{4}{*}{$10$}  & \multirow{1}{*}{$0.05$}  & $4.82910 - 0.34284 i$  & $4.813611 - 0.344125 i$\tabularnewline

\cline{2-5} 
 &  \multirow{1}{*}{$0.1$} & $ 4.68555 - 0.33564 i$  & $4.669033 - 0.336363 i$\tabularnewline

\cline{2-5} \cline{3-5}  
  & \multirow{1}{*}{$0.15$} & $4.56265 - 0.33042 i$ & $4.545331 - 0.330776 i$\tabularnewline
 
\cline{2-5}
 &  \multirow{1}{*}{$0.2$}  & $4.45571 - 0.32637 i$ & $4.437825 - 0.326469 i$ \tabularnewline
\cline{5-5} 
\hline 

\end{tabular}
\par\end{centering}
\caption{In this table we compare the fundamental QNM frequency ($n = 0$) of vector mode corresponding to gravitational perturbation of five dimensional Einstein-Gauss-Bonnet BH obtained using analytical and numerical methods for various choices of $\ell$ and $\alpha_{2}$.}
\label{vector_perturbation_qnm}
\end{table}
%%%%%%%%%%%%%%%%%%%%
%%%%%%%%%%%%%%%%%%%%
%%%%%%%%%%%%%%%%%%%%
%%%%%%%%%%%%%%%%%%%%

%%%%%%%%%%%%%%%%%%%%
%%%%%%%%%%%%%%%%%%%%
%%%%%%%%%%%%%%%%%%%%
%%%%%%%%%%%%%%%%%%%%
\begin{table}[h]
\begin{centering}
\begin{tabular}{|c|c|c|c|c|}
\hline 
$\ell$  &  $\alpha_{2}$ & Eikonal  & Numerical  \tabularnewline
\hline 
\hline 
\multirow{4}{*}{$2$}  & \multirow{1}{*}{$0.05$}  & $0.925604 - 0.351924 i
$  & $0.819891 - 0.331029 i$ \tabularnewline

\cline{2-5} 
 &  \multirow{1}{*}{$0.1$} & $0.871694 - 0.357668 i
$  & $0.758576 - 0.350335 i$  \tabularnewline

\cline{2-5} \cline{2-5}  
  & \multirow{1}{*}{$0.15$} & $0.830895 - 0.361760 i$  & $0.744297 - 0.354655 i$  \tabularnewline
 
\cline{2-5}
 &  \multirow{1}{*}{$0.2$}  & $ $   & $\text{instability}$  \tabularnewline
 
\cline{5-5} 
\hline 

\cline{5-5} 
\hline 
\multirow{4}{*}{$3$}  & \multirow{1}{*}{$0.05$}  & $1.38841 - 0.35192 i
$ & $1.334141 - 0.339535 i$ \tabularnewline

\cline{2-5} 
 &  \multirow{1}{*}{$0.1$} & $1.307541 - 0.357668 i$  & $1.263970 - 0.363322 i$ \tabularnewline

\cline{2-5} \cline{3-5}  
  & \multirow{1}{*}{$0.15$} &   & instability \tabularnewline
 
\cline{2-5}
 &  \multirow{1}{*}{$0.2$}  & $ $  & $\text{instability}$\tabularnewline
 
\cline{5-5} 
\hline 

\cline{5-5} 
\hline 
\multirow{4}{*}{$5$}  & \multirow{1}{*}{$0.05$}  & $2.31401 - 0.35192 i$  & $2.27226 - 0.351858 i$\tabularnewline

\cline{2-5} 
 &  \multirow{1}{*}{$0.1$} & $ 2.17924 - 0.35767 i$  & $2.13979 - 0.357834 i$\tabularnewline

\cline{2-5} \cline{3-5}  
  & \multirow{1}{*}{$0.15$} & $ $ & $\text{instability}$\tabularnewline
 
\cline{2-5}
 &  \multirow{1}{*}{$0.2$}  & $ $ & $\text{instability}
$ \tabularnewline
\cline{5-5} 
\hline 

\cline{5-5} 
\hline 
\multirow{4}{*}{$10$}  & \multirow{1}{*}{$0.05$}  & $ 4.62802 - 0.35192 i$  & $4.61087 - 0.353229 i$\tabularnewline

\cline{2-5} 
 &  \multirow{1}{*}{$0.1$} & $ 4.35847 - 0.35767 i$  & $4.34127 - 0.358732 i$\tabularnewline

\cline{2-5} \cline{3-5}  
  & \multirow{1}{*}{$0.15$} & $ $ & $\text{instability}$\tabularnewline
 
\cline{2-5}
 &  \multirow{1}{*}{$0.2$}  & $ $ & $\text{instability}$ \tabularnewline
\cline{5-5} 
\hline 

\end{tabular}
\par\end{centering}
\caption{The fundamental QNM frequencies ($n = 0$) associated with scalar mode of gravitational perturbation over the Einstein-Gauss-Bonnet BH in $d=5$ dimension have been compared between analytical and numerical results for various choices of $\ell$ and $\alpha_{2}$. The QNM frequencies of the scalar mode for higher $\alpha_{2}$ depicts instabilities, as evident from this table.}
\label{scalar_perturbation_qnm}
\end{table}
%%%%%%%%%%%%%%%%%%%%
%%%%%%%%%%%%%%%%%%%%
%%%%%%%%%%%%%%%%%%%%
%%%%%%%%%%%%%%%%%%%%

%%%%%%%%%%%%%%%%%%%%
%%%%%%%%%%%%%%%%%%%%
%%%%%%%%%%%%%%%%%%%%
%%%%%%%%%%%%%%%%%%%%
\begin{table}[h]
\begin{centering}
\begin{tabular}{|c|c|c|c|c|}
\hline 
$d$  &  $\ell$ & Eikonal  & Numerical  \tabularnewline
\hline 
\hline 
\multirow{4}{*}{$7$}  & \multirow{1}{*}{$2$}  & $0.486864 - 0.192459 i
$  & $0.432081 - 0.184924 i$ \tabularnewline

\cline{2-5} 
 &  \multirow{1}{*}{$3$} & $0.730297 - 0.192450 i
$  & $0.693884 - 0.190455 i $  \tabularnewline

\cline{2-5} \cline{2-5}  
  & \multirow{1}{*}{$5$} & $1.217162 - 0.192450 i$  & $1.195512 - 0.193074 i$  \tabularnewline
 
\cline{2-5}
 &  \multirow{1}{*}{$10$}  & $ 2.434321 - 0.192450 i $   & $2.424921 - 0.193879 i$  \tabularnewline
 
\cline{5-5}

\cline{2-5}
 &  \multirow{1}{*}{$20$}  & $ 4.868519 - 0.192450 i$   & $4.875322 - 0.193042 i$  \tabularnewline
 
\cline{5-5}
\hline 

\cline{5-5} 
\hline 
\multirow{4}{*}{$8$}  & \multirow{1}{*}{$2$}  & $0.581983 - 0.276059 i
$ & $0.483192 - 0.256851 i$ \tabularnewline

\cline{2-5} 
 &  \multirow{1}{*}{$3$} & $0.872975 - 0.276059 i$  & $0.807527 - 0.270096 i$ \tabularnewline

\cline{2-5} \cline{3-5}  
  & \multirow{1}{*}{$5$} & $1.454961 - 0.276059 i$  & $1.416090 - 0.276697 i$ \tabularnewline
 
\cline{2-5}
 &  \multirow{1}{*}{$10$}  & $2.909921 - 0.276059 i $  & $2.895231 - 0.278564 i$\tabularnewline
 
\cline{5-5} 

\cline{2-5}
 &  \multirow{1}{*}{$20$}  & $ 5.819832 - 0.276059 i$  & $5.830824 - 0.276956 i$\tabularnewline
 
\cline{5-5} 
\hline 

\end{tabular}
\par\end{centering}
\caption{We present a comparison of analytical and numerical results for the fundamental QNM frequencies ($n=0$) associated with the scalar mode of gravitational perturbation for pure Gauss-Bonnet BHs in $d=7$ and $d=8$ dimensions, for various values of $\ell$.}
\label{pure_lovelock_qnm}
\end{table}
%%%%%%%%%%%%%%%%%%%%
%%%%%%%%%%%%%%%%%%%%
%%%%%%%%%%%%%%%%%%%%
%%%%%%%%%%%%%%%%%%%%

%%%%%%%%%%%%%%%%%%%%%%%%%%%%%%%%%%%%%%%%%%%%%%%%%%%%%%%%%%%%%%%%%%%%%%%%%%%%%%%%%%%
%%%%%%%%%%%%%%%%%%%%%%%%%%%%%%%%%%%%%%%%%%%%%%%%%%%%%%%%%%%%%%%%%%%%%%%%%%%%%%%%%%%
%%%%%%%%%%%%%%%%%%%%%%%%%%%%%%%%%%%%%%%%%%%%%%%%%%%%%%%%%%%%%%%%%%%%%%%%%%%%%%%%%%%
\section{Conclusions}\label{sec:5}

In this paper, we have studied the eikonal QNMs associated with gravitational perturbations of static spherically symmetric BHs in the Lovelock theories of gravity (specifically, Einstein-Gauss-Bonnet gravity and pure Lovelock theories). In the case of Einstein-Gauss-Bonnet gravity, we have treated the higher order Gauss-Bonnet term as a perturbative correction to general relativity, since the equations of motion are known to be non-hyperbolic otherwise~\cite{papallo2015JHEP, brustein2018PRD}. The same also follows from the existence of stable null geodesic in the effective graviton metric for BHs in Einstein-Gauss-Bonnet gravity, with higher values of the Gauss-Bonnet coupling parameter $\alpha$, as we have depicted in this work. On the other hand, the null geodesics associated with BHs in pure Lovelock theories of gravity are always unstable and, hence, keep the field equations hyperbolic. Further, we have established through analytical means that the eikonal QNMs associated with gravitational perturbation are indeed linked with the unstable null geodesics of the static spherically symmetric BHs in Lovelock theories of gravity, though not with the physical metric, but with respect to the effective metrics. The same assertion has also been verified by numerical analysis, where the numerical values of the QNM frequencies in the eikonal limit match extremely well with the analytical expressions. 

This exercise has shown the intimate connection between the QNM frequencies associated with gravitational perturbation in a modified theory of gravity with its causal nature. For example, gravitational waves in Lovelock theories of gravity do not propagate along null lines and hence can be superluminal or subluminal, resulting into a possible issue of causality in these theories. The way out of this conundrum and to establish causality in Lovelock theories, it is necessary to work with an effective metric, rather than the physical metric, in which the gravitational waves propagate along null hypersurfaces. Thus, these effective graviton metrics, which differ between different polarizations of gravitational waves, are the geometry that a propagating gravitational wave `sees'. Thus, the causality of Lovelock theories hinges on the existence of such effective metric for the propagation of gravitational waves. In the present work, we have observed that the gravitational QNM frequencies are \emph{not} related to the properties of unstable null circular geodesics of the physical metric, but rather with these effective (graviton) metrics in the eikonal limit. Thereby relating causality in Lovelock theories of gravity with the QNM frequencies of gravitational perturbation in the eikonal limit. This is also in agreement with~\cite{glampedakis2019PRD}.

We have demonstrated the above explicitly for both Einstein-Gauss-Bonnet gravity as well as for pure Lovelock gravity theories, but can be generalized to any generic Lovelock polynomial in a straightforward manner. Also the instabilities in the QNM frequencies associated with scalar mode of gravitational perturbation for BHs in Einstein-Gauss-Bonnet gravity can be nicely mapped to the existence of stable null geodesics in the corresponding effective metric. The above correspondence between causality in Lovelock theories and BH QNM frequencies can also be verified by comparing the QNM frequencies derived analytically and numerically, which are in excellent agreement, as the previous section demonstrates. In summary, our results provide a connection between the eikonal QNMs associated with gravitational perturbation of BHs in Lovelock theories and the causal properties of the theory via the characteristic hypersurfaces and the \textit{bi-characteristic curves}~\cite{papallo2015JHEP}. This leads us to speculate whether the gravitational QNMs of BHs in the eikonal limit are related to the properties of the graviton-sphere in any higher curvature theories of gravity. In addition, if there exists non-minimal coupling between scalar as well as electromagnetic perturbation with curvature, the QNMs associated with scalar and electromagnetic perturbation will also be related to null geodesics of some effective metric and \emph{not} the physical BH spacetime. This, however, requires further analysis, which is beyond the scope of the present work.

On the observational side, in the current era of multi-messenger astronomy, gravitational lensing can be used to study the properties of the gravitonsphere, and our results suggest that it will indeed be possible to link such observations with the eikonal QNM frequencies. In general relativity, where the gravitational waves propagate along the null geodesics of the physical metric, observation of BH QNMs can be supplemented by BH image observations. More specifically, the real part of eikonal QNM frequencies of BHs in general relativity can be mapped to the size of the critical curve in the image, and the imaginary part can be mapped to the detailed photon ring structures~\cite{chen2023PLB}. If one observes that the properties of the BH shadow (which is effectively the photon ring) obtained from the remnant of a binary BH merger (possibly in the environment of an accretion disk, providing background illumination for the show to be visible) does not match with the QNM frequencies obtained from the GW coalescence signal, it will be a telltale signature of higher curvature corrections to general relativity. However, such observation is only possible using next-generation EHT (ngEHT)~\cite{blackburn2019arxiv} or future space-based Very-Long-Baseline Interferometry (VLBI) missions~\cite{johnson2020SA, haworth2019arxiv, gralla2020PRD}. Hence, the above connection between the QNM frequencies and causality in higher curvature theories can have interesting observational prospects in the not-so-distant future. 

%%%%%%%%%%%%%%%%%%%%%%%%%%%%%%%%%%%%%%%%%%%%%%%%%%%%%%%%%%%%%%%%%%%%%%%%%%%%%%%%%%%
%%%%%%%%%%%%%%%%%%%%%%%%%%%%%%%%%%%%%%%%%%%%%%%%%%%%%%%%%%%%%%%%%%%%%%%%%%%%%%%%%%%
%%%%%%%%%%%%%%%%%%%%%%%%%%%%%%%%%%%%%%%%%%%%%%%%%%%%%%%%%%%%%%%%%%%%%%%%%%%%%%%%%%%
\begin{acknowledgements}
AC acknowledges hospitality at IUCAA, Pune, where a part of the work was done. AC is financially supported by the Science and Engineering Research Board (SERB), Government of India through the National Post Doctoral Fellowship (File No. PDF/2023/000550). AKM would like to thank IACS Kolkata for hospitality, where part of this work was carried out. Research of SC is supported by MATRICS (MTR/2023/000049) and Core Research (CRG/2023/000934) Grants from SERB, ANRF, Government of India. SC also acknowledges the warm hospitality at the Albert-Einstein Institute, Potsdam, where a part of this work was done. The visit was supported by a Max-Planck-India mobility grant.
\end{acknowledgements}
%%%%%%%%%%%%%%%%%%%%%%%%%%%%%%%%%%%%%%%%%%%%%%%%%%%%%%%%%%%%%%%%%%%%%%%%%%%%%%%%%%%
%%%%%%%%%%%%%%%%%%%%%%%%%%%%%%%%%%%%%%%%%%%%%%%%%%%%%%%%%%%%%%%%%%%%%%%%%%%%%%%%%%%
%%%%%%%%%%%%%%%%%%%%%%%%%%%%%%%%%%%%%%%%%%%%%%%%%%%%%%%%%%%%%%%%%%%%%%%%%%%%%%%%%%%
\appendix
%%%%%%%%%%%%%%%%%%%%%%%%%%%%%%%%%%%%%%%%%%%%%%%%%%%%%%%%%%%%%%%%%%%%%%%%%%%%%%%%%%%
%%%%%%%%%%%%%%%%%%%%%%%%%%%%%%%%%%%%%%%%%%%%%%%%%%%%%%%%%%%%%%%%%%%%%%%%%%%%%%%%%%%
%%%%%%%%%%%%%%%%%%%%%%%%%%%%%%%%%%%%%%%%%%%%%%%%%%%%%%%%%%%%%%%%%%%%%%%%%%%%%%%%%%%
\section{Various Expressions}\label{AppA}

In this appendix we present the expressions for photon and graviton sphere, as well as the angular velocity and the Lyapunov exponent of the null circular geodesic of the photon and graviton metrics, expanded in terms of the Gauss-Bonnet coupling parameter $\alpha_{2}$. We obtain the following expressions,
%%%%%%%%%%%%%%%%%%%%%%%%%%%%%%%%%%%%%%%%%%%%%%%%%%%%%%%%%%%%%%
\begin{align}
r_c^A&=r_{c0}^A+r_{c1}^A \alpha_2+\mathcal{O}(\alpha_2^2)~,
\\
\Omega_c^A&=\Omega_{c0}^A+\Omega_{c1}^A \alpha_2+\mathcal{O}(\alpha_2^2)~,
\\
\lambda_c^A&=\lambda_{c0}^A+\lambda_{c1}^A \alpha_2+\lambda_{c2}^A \alpha_2^2+\mathcal{O}(\alpha_2^3)~,
\end{align}
%%%%%%%%%%%%%%%%%%%%%%%%%%%%%%%%%%%%%%%%%%%%%%%%%%%%%%%%%%%%%%
where,
%%%%%%%%%%%%%%%%%%%%%%%%%%%%%%%%%%%%%%%%%%%%%%%%%%%%%%%%%%%%%%
\begin{align}
r_{c0}^A&=\left(\frac{2}{(d-1)\mu}\right)^\frac{1}{3-d}~,
\label{eq:rc0}
\\
\Omega_{c0}^A&=\sqrt{\frac{d-3}{d-1}} \left(\frac{2}{(d-1) \mu }\right)^{\frac{1}{d-3}}~,
\label{eq:Oc0}
\\
\nonumber
\\
\lambda_{c0}^A&=\frac{ (d-3) \left(\frac{2}{(d-1) \mu }\right)^{\frac{1}{d-3}}}{\sqrt{d-1}}~,
\label{eq:lc0}
\\
\lambda_{c1}^A&=-\frac{(d-4) (d-3) (d-2) \alpha_2 \mu  \left(\frac{2}{(d-1) \mu }\right)^{\frac{d}{d-3}}}{2\sqrt{d-1}}~.
\label{eq:lc1}
\end{align}
%%%%%%%%%%%%%%%%%%%%%%%%%%%%%%%%%%%%%%%%%%%%%%%%%%%%%%%%%%%%%%
We would like to emphasize that the leading order terms as presented in Eqs.~(\ref{eq:rc0}-\ref{eq:lc1}) are identical for the photon and graviton spheres. The leading order differences have been depicted in the main text. 
\bibliography{ref} 
\end{document}